\documentclass[showpacs,preprintnumbers,amsmath,aps,nofootinbib,amssymb,superscriptaddress,twocolumn]{revtex4}
\usepackage{float}
\usepackage[caption=false,justification=justified]{subfig}  
\usepackage{graphicx}
\usepackage[dvipsnames]{xcolor}
\usepackage[english]{babel}
\usepackage{bm}
\usepackage[font=scriptsize]{caption}
\usepackage{ragged2e}
\usepackage[font=small,labelfont=bf,justification=raggedright]{caption}
\usepackage[utf8]{inputenc}
\usepackage[tikz]{bclogo}
\usepackage[framemethod=tikz]{mdframed}

\newsavebox{\measurebox}
\usepackage[colorlinks=false]{hyperref}
\begin{document}
\baselineskip 16pt
\title{Noncommutative effective LQC: inclusion of potential term.}

\author{Luis Rey D\'iaz-Barr\'on}
\email{lrdiaz@ipn.mx}
\affiliation{Unidad Profesional Interdisciplinaria de Ingenier\'ia
Campus Guana\-jua\-to del Instituto Polit\'ecnico Nacional.\\
Av. Mineral de Valenciana No. 200, Col. Fraccionamiento Industrial Puerto Interior, C.P. 36275, Silao de la Victoria, Guana\-jua\.to,
M\'exico.}

\author{Abraham Espinoza-Garc\'ia}
\email{aespinoza@ipn.mx}
\affiliation{Unidad Profesional Interdisciplinaria de Ingenier\'ia
Campus Guana\-jua\-to del Instituto Polit\'ecnico Nacional.\\
Av. Mineral de Valenciana No. 200, Col. Fraccionamiento Industrial Puerto Interior, C.P. 36275, Silao de la Victoria, Guana\-jua\.to,
M\'exico.}

\author{S. P\'erez-Pay\'an}
\email{saperezp@ipn.mx}
\affiliation{Unidad Profesional Interdisciplinaria de Ingenier\'ia
Campus Guana\-jua\-to del Instituto Polit\'ecnico Nacional.\\
Av. Mineral de Valenciana No. 200, Col. Fraccionamiento Industrial Puerto Interior, C.P. 36275, Silao de la Victoria, Guana\-jua\.to,
M\'exico.}

\author{J. Socorro}
\email{socorro@fisica.ugto.mx}
\affiliation{Departamento de
F\'{\i}sica, DCeI, Universidad de Guanajuato-Campus Le\'on, C.P.
37150, Le\'on, Guanajuato, M\'exico.}
	
\begin{abstract}
We construct and study a simple noncommutative scheme (theta-deformation) for the effective Loop Quantum Cosmology of the flat Friedmann-Lema\^itre-Robertson-Walker model in the presence of a homogeneous scalar field $\phi$ with a potential $\mathcal{V}(\phi)=\frac{1}{2}m^2\phi^2$. We first conduct a simple analysis from the corresponding Hamilton equations of motion considering a generic term $\mathcal V(\phi)$. It is observed that the characteristic Big Bounce of Loop Quantum Cosmology is preserved under such noncommutative extension. When  specializing to the quadratic case, numerical solutions to the corresponding Hamilton equations exhibiting an early inflationary epoch with a sufficiently large number of e-foldings are found. It is concluded that, in this noncommutative setup,  
solutions exist which are in the overall compatible with the early universe predicted by standard (effective) Loop Quantum Cosmology (i.e. a bouncing and inflationary early  universe). The issue of the genericness of a sufficiently long inflationary period on the space of solutions in this noncommutative construct remains to be addressed.
\end{abstract}
	
	
\maketitle

\section{Introduction}
The somewhat recent investigations \cite{connes-douglas}, \cite{seiberg-witten} resulted in the resurrection of the noncommutativity paradigm, which had remained burried for roughly five decades. The original idea can be traced back to the work of Snyder \cite{snyder}, where noncommutativity in fixed Minkowski spacetime is proposed as a mechanism to regularize quantum field theory. On the other hand, the theoretical evidence provided by leading approaches to quantum gravity (e.g. string theory, loop quantum gravity (LQG) ---a standard monograph is \cite{thiemann}) that the fundamental structure of spacetime is expected to incorporate some sort of discreteness, has motivated the use of the noncommutativity paradigm to approximately model such a granular picture, and study its consequences (see the reviews \cite{nekrasov},\cite{szabo} for a string theory community perspective of noncommutative field theory). Although most of the noncommutative constructs have relied upon a fixed flat spacetime background, some attempts to incorporate noncommutativity directly into the full gravitational field have been performed (see, for instance, \cite{mofat}-\cite{kober}). The inherent complexity of these proposals have led theoreticians to consider toy models in which the gravitational field can be described by a finite number of degrees of freedom in order to extract, in a simple way, some of the physical consequences of noncommutativity. In the seminal investigation \cite{compean3} the authors implemented a star product on the Wheeler-DeWitt equation  of the Kantowski-Sachs anisotropic model in order to study the effects of an assumed noncommutativity (theta deformation) in the corresponding quantum minisuperspace. On the other hand, in \cite{barbosa}, the authors considered the effects of noncommutativity already at the classical configuration manifold of the Kantowski-Sachs spacetime. These two investigations spurred noncommutative extensions of relevant cosmological and astrophysical scenarios (e.g. \cite{wally-1}-\cite{Sinuhe_2}), at both classical and quantum levels. 

The present research work is a follow up to \cite{hindawi} and \cite{ijmpd}, in which a canonical noncommutativity at the effective scheme of the loop quantum cosmology (LQC) of the flat Friedmann-Lema\^itre-Robertson-Walker (FLRW) model in the presence of a \textit{free} homogeneous scalar field was considered. One of the main motivations of the work \cite{hindawi} was to study to what extend a simple theta deformation at the so called effective scheme could prompt radical modifications of key features of the LQC paradigm (e.g. singularity resolution through a Big Bounce). It was concluded that a triad-scalar field momentum theta-deformation does not modify the scalar field energy density, furthermore, such noncommutativity allows to maintain the single Big Bounce (associated to the maximum value $\rho_c$ of the energy density) of the standard LQC paradigm. Additionally, in \cite{ijmpd} it was observed that as a result of a configuration  sector theta-deformation the \textit{a priori} free scalar field $\phi$ acquires an effective potential, and the question of whether this could prompt an inflationary phase with enough e-foldings was addressed. The answer was in the negative. On the other hand, it is well known that the inclusion of a homogeneous scalar field with quadratic potential in the effective LQC of the flat FLRW results in a period of inflation with enough e-foldings, and that this is a rather generic feature of the solutions: the \textit{a priori probability} of picking a solution which generates an inflationary period with at least 68 e-foldings is very close to one \cite{sloan}. Therefore, the (effective) LQC paradigm not only resolves the classical initial singularity, but it is also capable of reproducing the inflation mechanism rather generically. The present investigation aims at establishing whether inflationary solutions exist which do not depart radically from the early universe depicted by standard effective LQC (for the case of a quadratic potential term), when implementing a theta-deformation in the momentum sector of phase space.

The manuscript is organized as follows. Section II is devoted to the construction and study of a noncommutative extension of the standard case (which is briefly reviewed in this Introduction), but considering a generic potential term $\mathcal{V}(\phi)$. We undertake a simple analysis via the Hamiltonian constraint and the equations of motion, which enables us to single out a particular representation of the considered noncommutativity. Next, in section III, we specialize the discussion of section II to a quadratic potential. Here, numerical solutions which exhibit a sufficiently long early inflationary epoch are presented. Section IV is devoted to make some remarks and state some conclusions.

In the following we briefly review the effective LQC paradigm. 
%
In the LQC literature, it is well known (see, for instance, the reviews \cite{bojowald-lrr}, \cite{lqc-status-report}, and references therein) that in the case of the flat FLRW cosmology in the presence of a scalar field $\phi$ with potential term
\begin{equation}
\mathcal{V}(\phi)=\frac{1}{2} m^2\phi^2,\label{qpotential}
\end{equation}
the effective Hamiltonian takes the form
\begin{equation}
\mathcal{H}_{eff}=-\frac{3}{8\pi G\gamma^{2}\lambda^{2}}\sin^{2}(\lambda\beta)V+\frac{(p_{\phi})^2}{2V}+\frac{1}{2}m^2\phi^2V.\label{ham-const-lqc}
\end{equation}
The associated Hamilton equations are
\begin{eqnarray}
\label{eqs-motion-lqc-beta}\dot{\beta}&=&-\frac{3}{\gamma\lambda^2}\sin^{2}(\lambda\beta)+4\pi G\gamma m^{2}\phi^{2},\\ 
\label{eqs-motion-lqc-fi}\dot{\phi}&=&\frac{p_{\phi}}{V},\\
\label{eqs-motion-lqc-v}\dot{V}&=&\frac{3}{\gamma\lambda}V\sin(\lambda\beta)\cos(\lambda\beta),\\
\label{eqs-motion-lqc-pfi}\dot{p}_{\phi}&=&-m^{2}V\phi. 
\end{eqnarray}
This so called effective scheme is based on a geometrical formulation of quantum mechanics (QM) \cite{ashtekar-gfqm}. That this semiclassical approximation reproduces remarkably well the quantum behavior given by full LQC has been established via numerical simulations in \cite{diener1}. 

With the help of the Hamiltonian constraint, we readily note: 
\begin{itemize}
\item The equation of motion for $\beta$ indicates that this dynamical variable is a decreasing function of time. In exactly soluble LQC, it is shown that $\beta$ takes values in the interval $\left(0,\frac{\pi}{\lambda}\right)$\cite{robustness} (the interval $\left[\left.-\frac{\pi}{2\lambda},\frac{\pi}{2\lambda}\right)\right.$ is also employed in this effective scheme -- see for instance \cite{corichi-tatiana}). 
\item Equation for $V$ implies that a bounce takes place at $t=t_c$ with $\beta(t_c)=\frac{\pi}{2\lambda}$, and that this bounce corresponds to the maximum of the energy density: $\rho(t_c)=\frac{1}{2}\dot{\phi}^2(t_c)+\frac{m^2}{2}\phi^2(t_c)=\frac{3}{8\pi G\gamma^{2}\lambda^{2}}\sin^{2}(\lambda\beta(t_c))=\frac{3}{8\pi G\gamma^{2}\lambda^{2}}$ (where in arriving at the second equality the equation of motion for $\phi$ and the Hamiltonian constraint were used). In the following, we denote by $\rho_c$ the maximum value attained by the energy density function. For definiteness, the bounce is naturally set at $t_c=0$. 
\item Since $\beta$ presents a monotonic behavior, such bounce is always featured and is achieved exactly once. Therefore, in this effective scheme, the Big Bounce is generic. This singularity resolution via a Big Bounce is a key feature of (effective) LQC.
\end{itemize}

\begin{figure*}[htp]
\centering

   \subfloat[ ]{\label{rev0}
      \includegraphics[width=.25\textwidth]{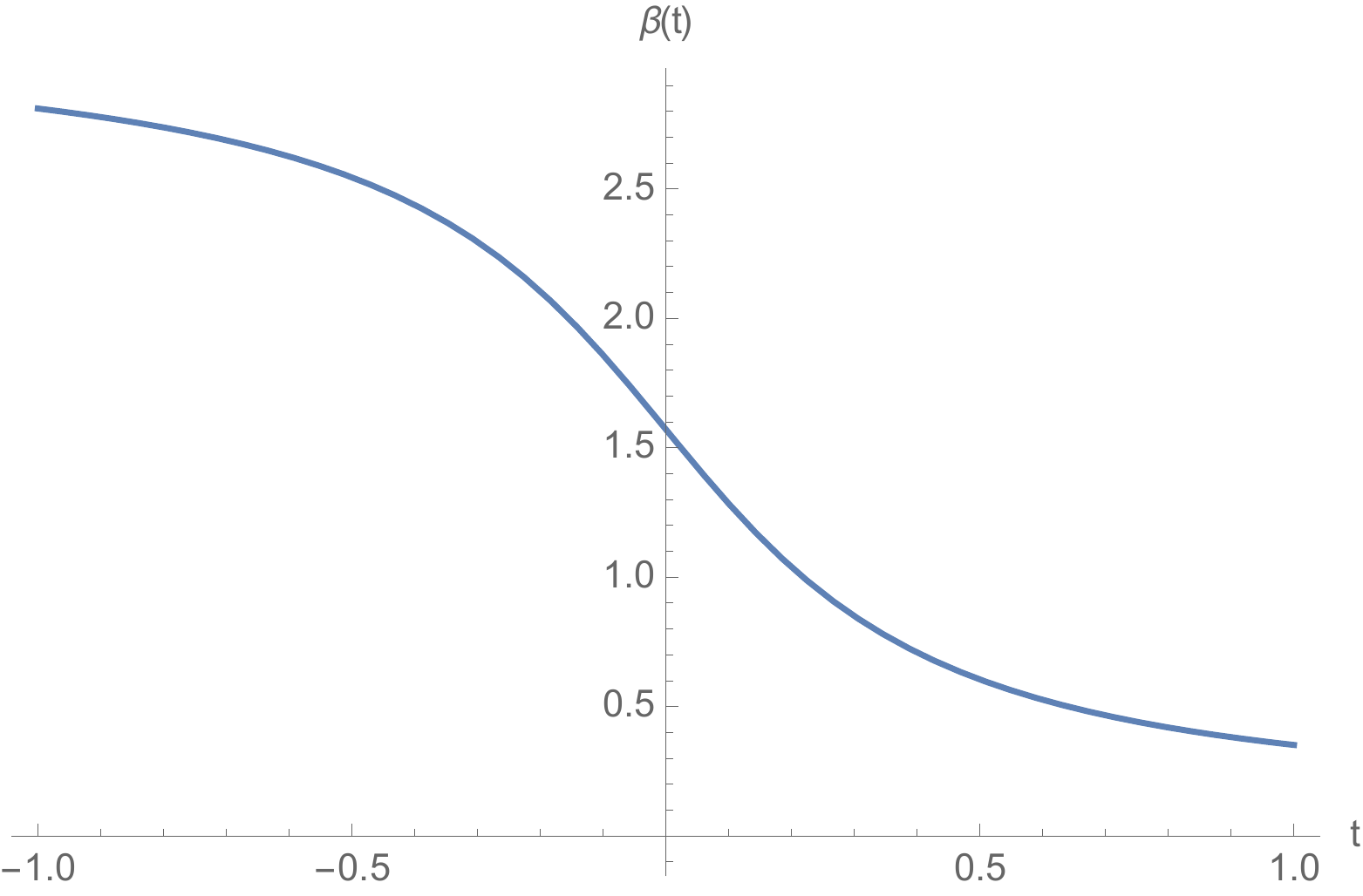}}
~
   \subfloat[ ]{\label{rev_sol1}
      \includegraphics[width=.25\textwidth]{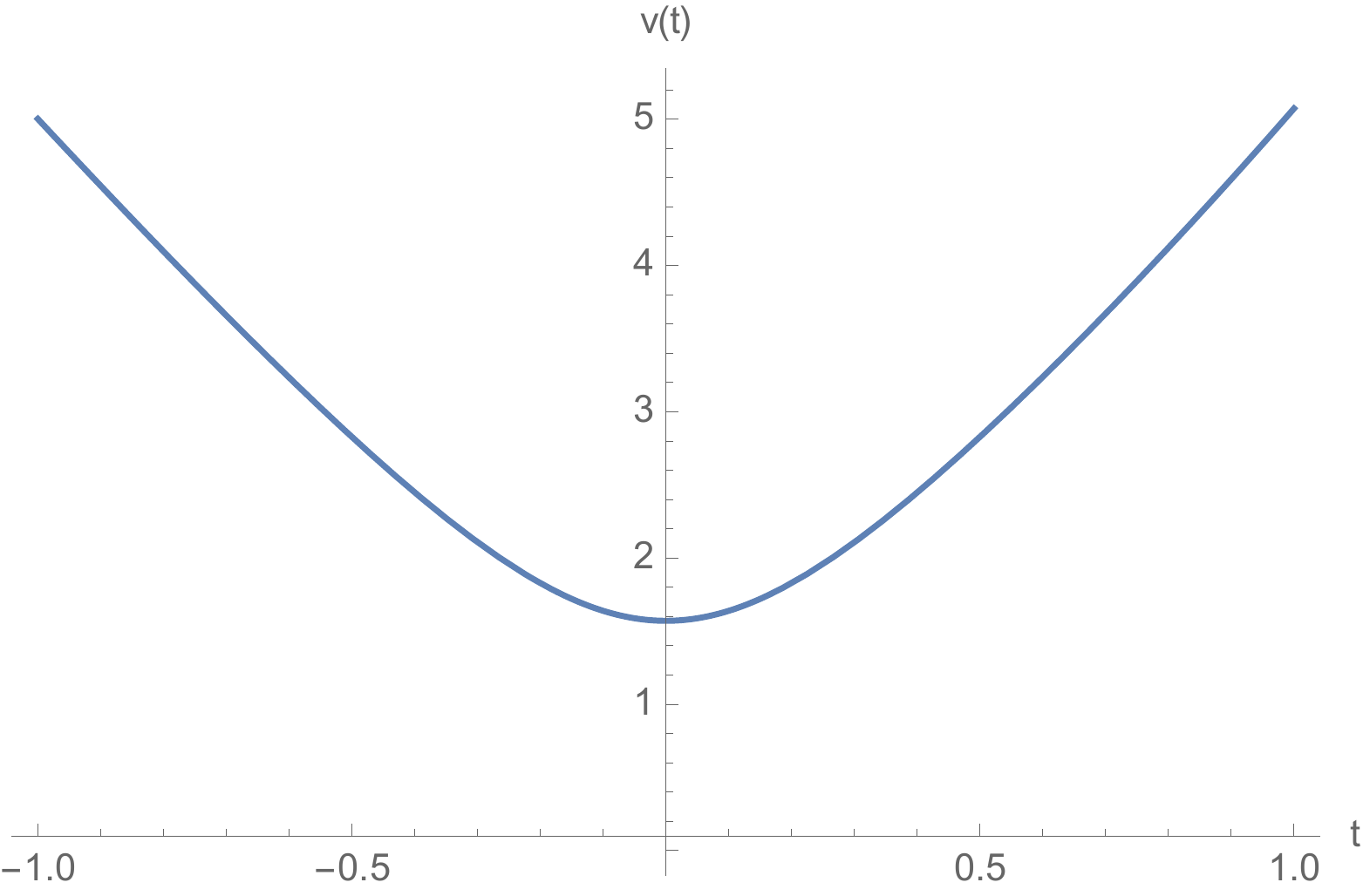}}

   \subfloat[ ]{\label{rev_so2}
      \includegraphics[width=.25\textwidth]{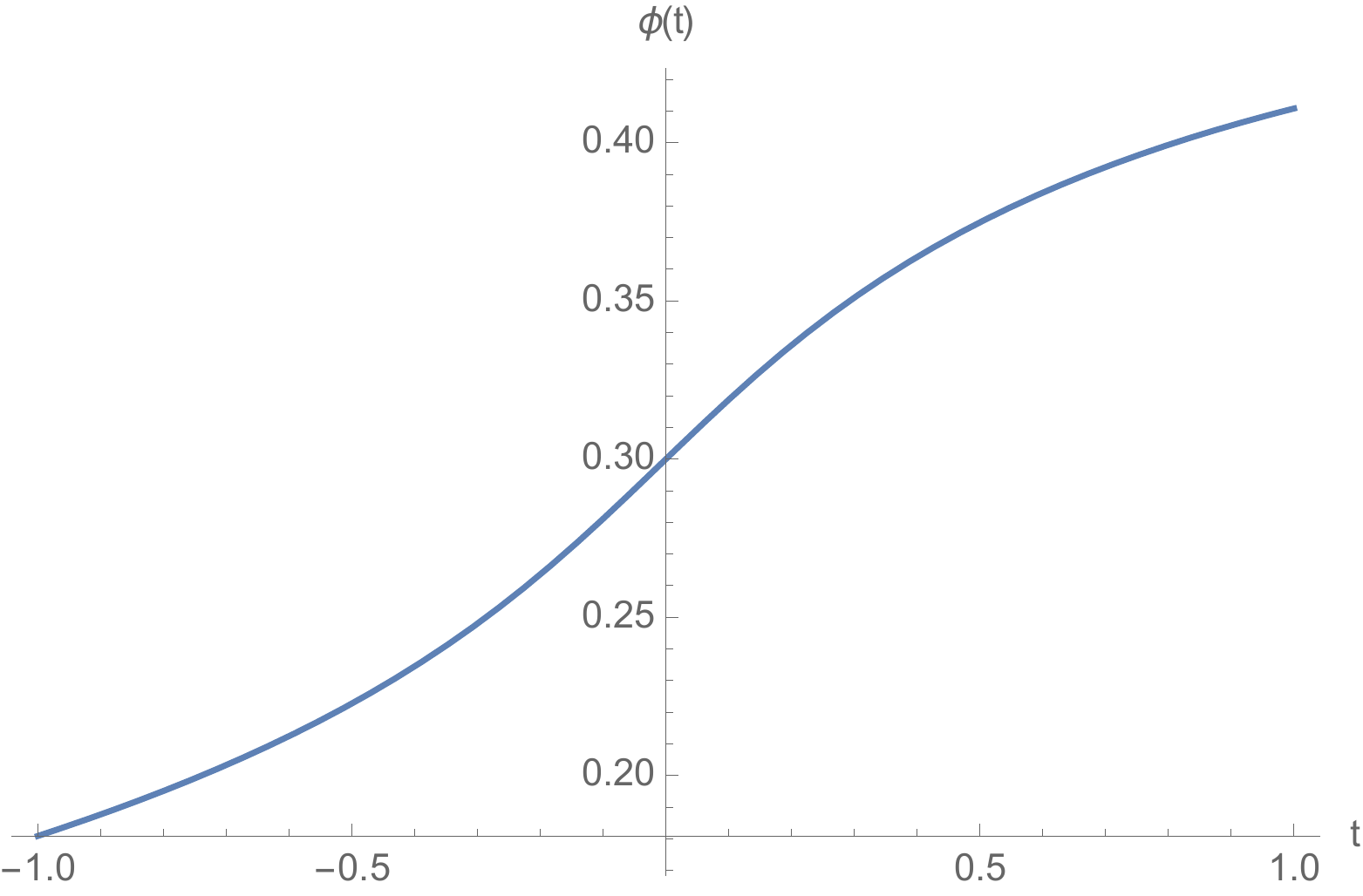}}
~
   \subfloat[ ]{\label{rev_sol3}
      \includegraphics[width=.25\textwidth]{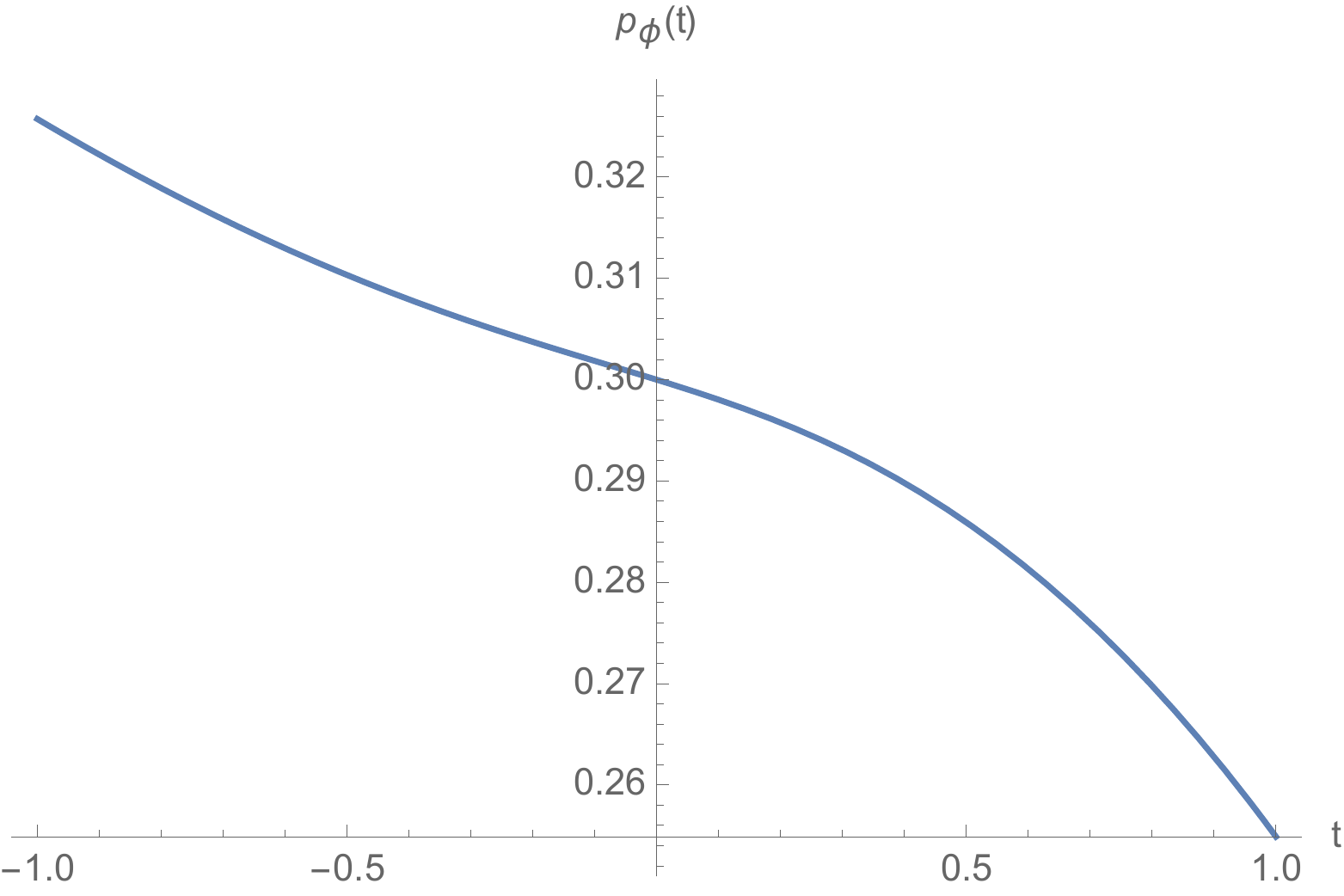}}
~
   \subfloat[ ]{\label{rev_sol4}
      \includegraphics[width=.25\textwidth]{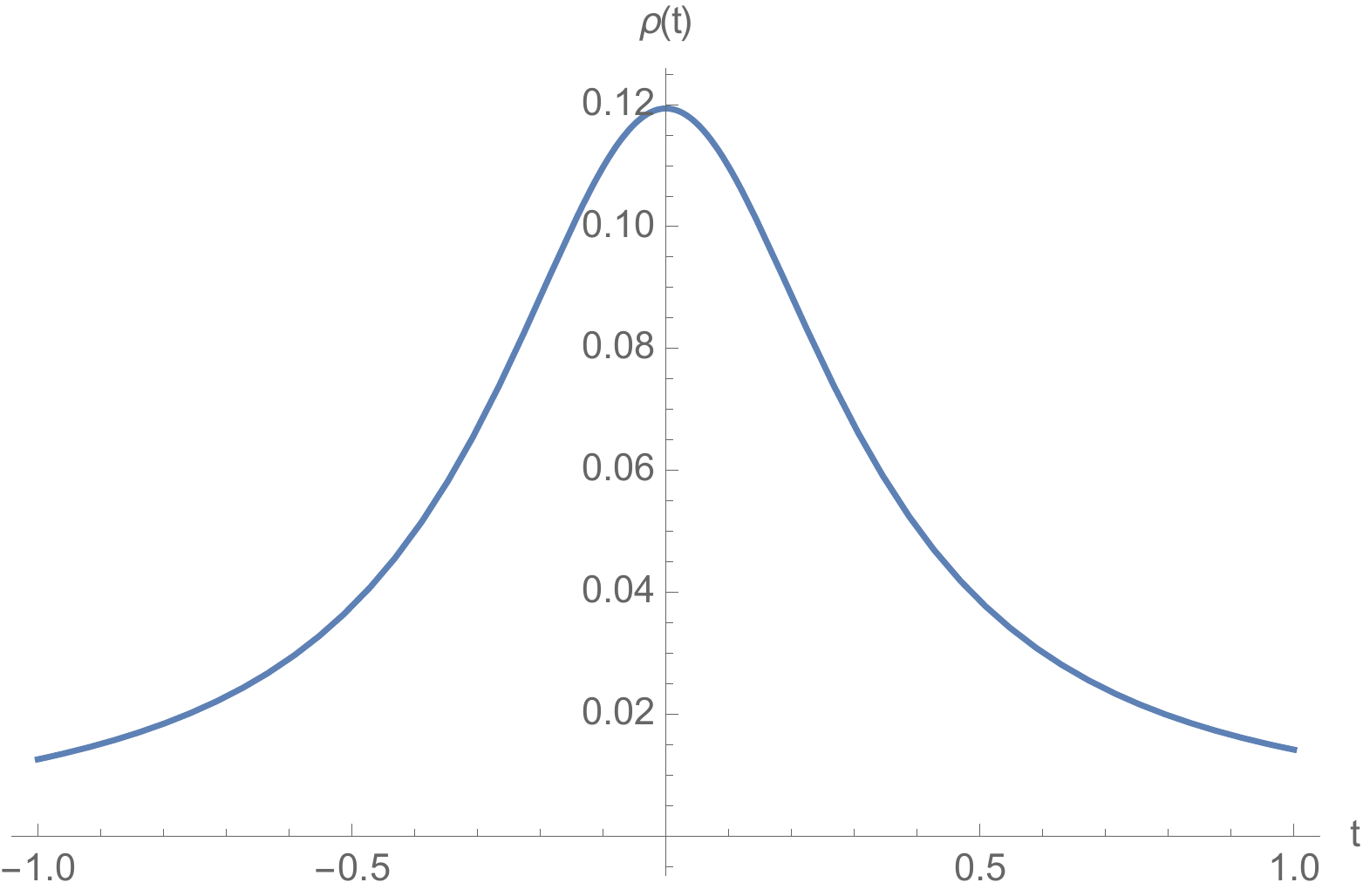}}

\caption{Numerical solutions to (\ref{eqs-motion-lqc-beta})-(\ref{eqs-motion-lqc-pfi}). (a) $\beta(t)$. (b) volume function, $V(t)$. (c) scalar field, $\phi(t)$. (d) momentum $p_\phi(t)$ canonically conjugated to the scalar field; (e) energy density, $\rho(t)=\dot\phi^2/2+m^2\phi^2/2$. The bounce was set to take place at $t_c=0$. The following initial conditions and values were employed: $v(0)=\pi/2\lambda$, $\beta (0)=\pi/2\lambda$, $\phi (0)=0.3$, $p_\phi(0)=0.3$, an the constants $b=1$, $a=-1+4\pi$, $m=.2$ } 
\end{figure*}

\begin{figure}[b]
\begin{center}
\includegraphics[width=3in]{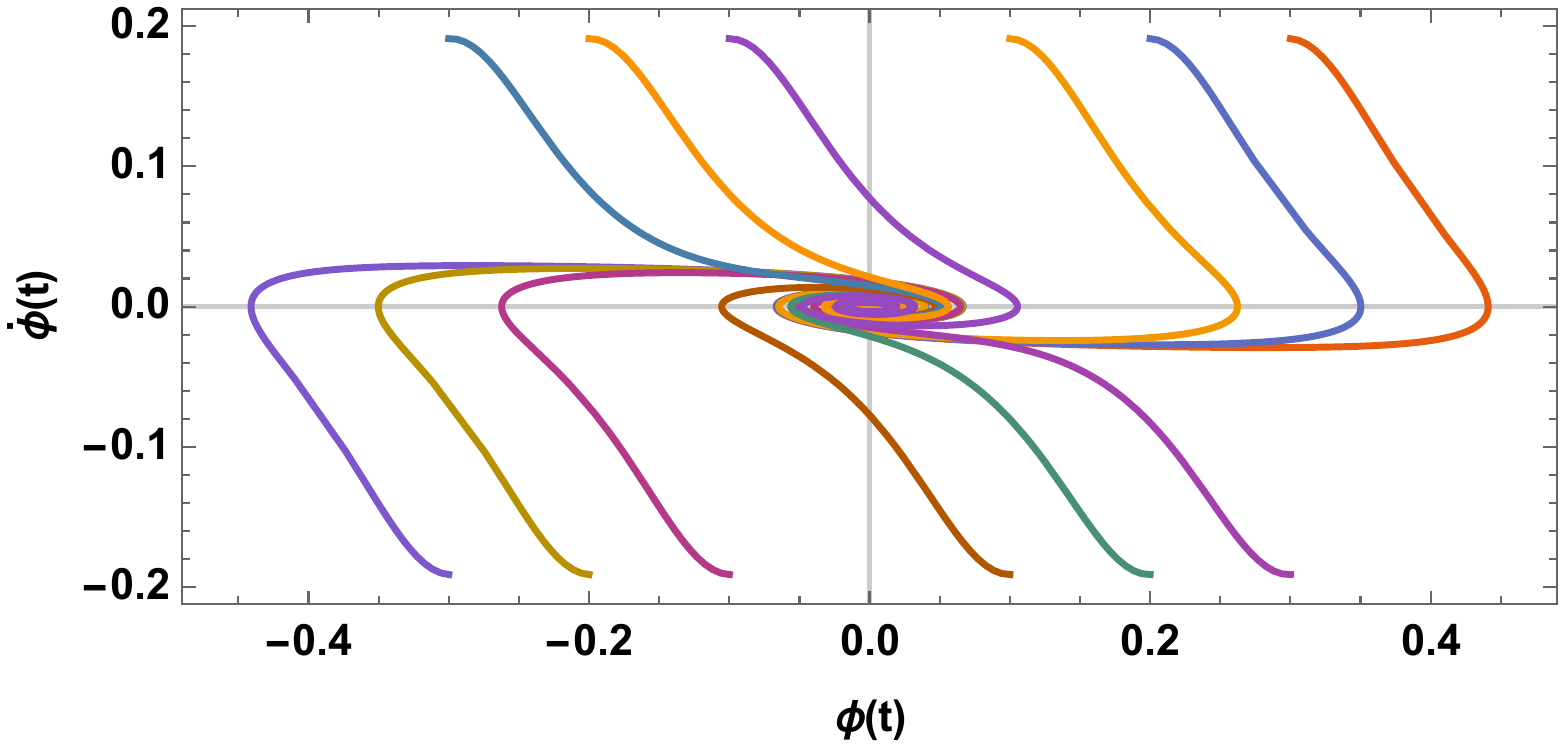}
\caption{Phase portrait of the scalar field $\phi$. The attractor is located at $\pm m/\sqrt {12\pi}$}
\label{fig:Atrsol1}
\end{center}
\end{figure}

For completeness, we obtained some numerical solutions to the equations of motion. A particular set of which is shown in Figs. \ref{rev0}-\ref{rev_sol4}. We also reproduce, in Fig. \ref{fig:Atrsol1}, the familiar phase portrait (compare with, e.g. \cite{mukhanov}, \cite{agullo-corichi}) for the scalar field $\phi$ by explicitly evaluating some of the numerical solutions obtained.
%
%
Now, the corresponding Friedmann equation and Klein-Gordon equation are given by ($H=\frac{\dot{a}}{a}=\frac{\dot{V}}{3V}$ is the Hubble parameter):
\begin{equation}
\label{FKGe}
H^2=\frac{8\pi G}{3}\rho\left(1-\frac{\rho}{\rho_c}\right),\qquad \ddot\phi+3H\dot\phi+m^2\phi=0.
\end{equation}
Note that the singularity resolution is incorporated in this modified Friedmann equation in a rather simple way: $\dot{a}=0$ at $\rho=\rho_c$. 
\section{Noncommutative effective LQC of the flat FLRW with scalar field $\phi$ with generic potential term $\mathcal{V}(\phi)$}\label{section2}
We first review the ideas and methods related to the noncommutativity paradigm as implemented in the minisuperspace approximation.  

Recall (see, e.g. \cite{choquet}) that the phase space $\Gamma=T^\star\mathcal{Q}$ (the cotangent bundle of $\mathcal{Q}$), where $\mathcal{Q}$ is the configuration space of a mechanical system, is a symplectic manifold: a $2n$-dimensional manifold endowed with a closed ($d\omega=0$) and invertible ($\mathrm{rank}(\omega)=2n$) two-form $\omega=\omega_{\alpha\beta}dx^\alpha\wedge dx^\beta$, called the symplectic structure (in this particular case of the cotangent bundle,  $\omega$ is in fact exact: $\omega=d\vartheta$, $\vartheta$ being the so called symplectic potential).

The symplectic structure provides an isomorphism $\omega^\flat:T_p\mathcal{Q}\rightarrow T_p^\star\mathcal{Q}$, for each $p\in\mathcal Q$, by $\omega^\flat(\mathsf X)=\omega(\mathsf X,\cdot)$; with inverse $\omega^\sharp:T_p^\star{\mathcal Q}\rightarrow T_p\mathcal Q$, for each $p\in\mathcal Q$, given by $\omega^\sharp(\eta)=\mathsf{P}(\eta,\cdot)$, where  $\mathsf{P}=\omega^{\alpha\beta}\partial_\alpha\wedge\partial_\beta$ is the Poisson structure (an antisymmetric bivector field on $T^\star\mathcal Q$ satisfying $\omega^{\alpha\beta}\omega_{\beta\gamma}=\delta^{\alpha}_{\gamma}$). 
As a consequence, one can write the equations of motion 
\begin{equation}
\dot{q}^a=\partial_{p_a}H,\quad\dot{p}_a=-\partial_{q^a}H \label{local-ham-sys}
\end{equation}
of a mechanical system with Hamiltonian function $H$ in a geometric and coordinate-independent way:
\begin{equation}
\dot{\mathsf{x}}=-\omega^\sharp(dH),\label{ham-sys}
\end{equation}
where $\mathsf{x}=(x^1,...,x^{2n})$ are the phase space dynamical variables, $dH=\partial_\alpha dx^\alpha$, and  $\omega^\sharp(dH)=-\mathsf{X}_H$ the associated vector field to $dH$: $-\mathsf{X}_H=\mathsf{P}(dH,\cdot)\leftrightarrow\omega(-\mathsf X_H,\cdot)=dH$.

In canonical coordinates $(x^1,...,x^{2n})=(q^1,q^2,...,q^n,p_1,p_2,...,p_n)$ (i.e. in a Darboux chart), the symplectic structure acquires the usual form ($a=1,...,n$)
\begin{equation}
\omega=J_{\alpha\beta}dx^{\alpha}\wedge dx^{\beta}=dx^{n+a}\wedge dx^{a}=dp_{a}\wedge dq^{a},
\end{equation}
that is,  
\begin{equation}
J_{\alpha\beta}=\begin{pmatrix}
0_{n} &-I_{n}\\
I_{n} &0_{n}
\end{pmatrix}.
\end{equation}
We therefore have (in canonical coordinates)
\begin{equation}
\mathsf{P}=J^{\alpha\beta}\partial_{\alpha}\wedge\partial_{\beta}=\partial_a\wedge\partial_{n+a}=\partial_{q^{a}}\wedge\partial_{p_{a}},
\end{equation}
and so, 
\begin{equation}
J^{\alpha\beta}=\begin{pmatrix}
0_{n} &I_{n}\\
-I_{n} &0_{n}
\end{pmatrix}.\label{flat-structure}
\end{equation}
(We reserve the notation $J_{\alpha\beta}$, $J^{\alpha\beta}$ for the components of the symplectic and Poisson structures when expressed in a Darboux charts ---i.e. in canonical coordinates.) 

Recall also that Poisson brackets can be defined by the Poisson structure or the symplectic structure:
\begin{equation}
-\omega(\mathsf X_f,\mathsf X_g)=:\{f,g\}:=\mathsf{P}(df,dg)
\end{equation}
hence, in canonical coordinates, we arrive at the familiar form (sum over $a$):
\begin{equation}
\{f,g\}=\partial_{q^a}f\partial_{p_a}g-\partial_{q^a}g\partial_{p_a}f.
\end{equation}
In particular, for the canonical coordinates $(x^1,x^2,...,x^{2n})$, we have
\begin{equation}
\{q^{a},p_{b}\}=\delta^{a}_{b},\qquad\{q^{a},q^{b}\}=\{p_{a},p_{b}\}=0.
\label{c-flat relations}
\end{equation}
The so called correspondence principle leads to the canonical commutation relations
\begin{equation}
[\hat{q}^i,\hat{p}_j]=i\hbar\delta^i_j,\quad[\hat{q}^i,\hat{q}^j]=[\hat{p}_i,\hat{p}_j]=0\label{cancommrel}
\end{equation}
Now, leading approaches to quantum gravity (e.g. Loop Quantum Gravity) point to a discrete picture of spacetime at distances comparable to the Planck length $\ell_{p}\sim10^{-35}\mathrm{cm}$: i.e. in a full quantum theory of the gravitational field, a quantization of spacetime itself could be in order. A possible way to account for such discrete picture is provided by an uncertainty relation such as
\begin{equation}
[q^{i},q^{j}]=i\theta^{ij}.
\end{equation}
The noncommutativity paradigm which we consider here is closely related to \textit{Quantum Mechanics in Phase Space} (QMPS), which, as its name suggests, provides an alternate way to consistently formulate Quantum Mechanics in \textit{classical} phase space (an introductory account of QMPS can be found in \cite{curtright}). In essence, in QMPS, the canonical commutation relations (\ref{cancommrel}) are realized  by means of a \textit{deformation} of the usual pointwise multiplication of phase space functions: $f\cdot g\mapsto f\star g$. This \textit{star-product} ($\star$-product) can be thought of as a ``power series'' of $\frac{i\hbar}{2}\mathsf{P}(df,dg)$ (where $\mathsf{P}$ is the Poisson structure of classical phase space) in which the ``zeroth power'' is the usual commutative product $f\cdot g$. We can therefore obtain standard Classical Mechanics by simply taking $\hbar\to0$. By considering the (classical) commutator $[f,g]_\star=f\star g-g\star f$, the usual canonical commutation relations (\ref{cancommrel}) are realized in \textit{classical} phase space $\Gamma$. Central to the scheme of QMPS is the so called \textit{Wigner function}, a quasi-probability distribution in phase space which was originally introduced in the context of Classical Statistical Mechanics in order to account for quantum corrections to the classical Liouville measure. In QMPS, probability amplitudes are calculated with the help of the Wigner function.

The connection with canonical Quantum Mechanics is given by working in the Weyl representation (which is unitary equivalent to the standard Schr\"odinger representation), where the canonical commutation relations (\ref{cancommrel}) acquires an ``exponentiated form''; this representation is encoded in a quantization prescription, the so called Weyl map $\mathcal W$, which associates a (symmetric ordered) quantum operator $\hat{f}$ to a phase space function $f$. The Weyl map $\mathcal W$ is invertible ($\mathcal{W}^{-1}(\hat{f})$ is called the Weyl symbol of $\hat{f}$, or the Wigner transform of $\hat{f}$). The Wigner function maps (up to multiplicative factors of $\hbar$) to the density operator $\hat{\rho}$ under the Weyl transform (i.e. the Wigner function is the Wigner transform of the density operator). The $\star$-product is the image of the operator product (in the Weyl representation) under the Wigner transform: $f\star g=\mathcal{W}^{-1}(\mathcal{W}(f)\mathcal{W}(g))$.

For the case of the canonical algebra (\ref{cancommrel}) we have
\begin{eqnarray}
f\star g&=&\exp\left(\frac{i\hbar}{2}\mathsf P(df,dg)\right)
=f(x)\exp{\left(\frac{i\hbar}{2}\partial_{\mu}J^{\mu\nu}\partial_{\nu}\right)}g(x)\nonumber\\
&=&fg+\frac{1}{2}\{f,g\}+\mathcal{O}(\hbar^{2})
\label{moyal}
\end{eqnarray}
where in the expansion, terms of the form $\mathsf{P}^{r}(df,dg)=J^{\mu_{1}\nu_{1}}\cdot\cdot\cdot J^{\mu_{r}\nu_{r}}\left(\partial_{\mu_{1}...\mu_{r}}f\right)\left(\partial_{\nu_{1}...\nu_{r}}g\right)$ (the ``rth power'' of the Poisson bracket of $f$ and $g$) are to be considered. This is the Moyal $\star$-product \cite{moyal}, it replaces the ordinary pointwise multiplication in the algebra of functions defined in phase space. The Moyal $\star$-bracket $[f,g]_\star:=f\star g-g\star f$ is thus responsible for the realization of the canonical algebra (\ref{cancommrel}). 
This product, together with the Wigner function, is the cornerstone of QMPS.

The noncommutativity paradigm employs some of the ideas and methods of QMPS (most notably, the $\star$-product), and can therefore be viewed from the perspective of the QMPS scheme: given more general commutation relations, e.g.
\begin{equation}  
\left[\hat{q}^a,\hat{p}_b\right]=i\hbar\delta^a_b,\quad[\hat{q}^a,\hat{q}^b]=i\theta^{ab},\quad[\hat{p}_a,\hat{p}_b]=i\kappa_{ab}, \label{theta-def}
\end{equation}
construct the corresponding $\star$-product by accordingly considering an appropriate modification $\tilde{\mathsf{P}}$ to $\mathsf{P}$. This  $\tilde{\mathsf{P}}$ will be such that the classical Poisson relations corresponding to the above commutation relations are fulfilled. Once the $\star$-product is constructed, the advancement of a  Noncommutative Quantum Mechanics in Phase Space (NCQMPS) can be pursued. A related point of view is to consider  (\ref{theta-def}) as an extension to standard Quantum Mechanics, resulting in a more noncommutative framework than the original standard one. The implementation of such additional noncommutativity is achieved by implementing a $\star$-product on \textit{quantum} phase space. This point of view results in a framework known as Noncommutative Quantum Mechanics (NCQM) \cite{mezin}.

The introduction of the modification $\tilde{\mathsf{P}}$ can, at the classical level, be interpreted as a manifestation of the phase space coordinates not being canonical, which in turn would imply that a new phase space $\tilde\Gamma$ is being considered (a ``deformation'' to the original one, $\Gamma$). An interesting point of view is the assumption that this new $\tilde\Gamma$ \textit{is the true phase space of the mechanical system} under study, and that it was initially overlooked due to the noncommutative scale being unnoticed. This phase space $\tilde\Gamma$ would therefore carry (noncommutative) quantum corrections in a purely classical setup. In this interpretation, the classical equations of motion would not be in canonical form in the original (now non-canonical) variables $y^\alpha$, but the Hamiltonian function would remain unchanged. This deviation from canonicity of the equations of motion (\textit{noncommutative equations of motion}) would yield such quantum corrections. Due to the Darboux theorem, one would be able to find a new chart with local coordinates $x^\alpha$ ($y=T(x)$) in which $\tilde{\mathsf{P}}$ acquires the canonical form (\ref{flat-structure}) (so that the classical equations of motion would be in canonical form), but the Hamiltonian function would get modified ($H(y)\mapsto H(x)=H(T(x))$): the quantum corrections from noncommutativity would in this case descend to the Hamiltonian function (this point of view was adopted in Ref. \cite{barbosa} in order to present a noncommutative classical cosmology of the Kantowski-Sachs model). In our investigations, we follow this latter \textit{deformed Hamiltonian} (or \textit{noncommutative Hamiltonian}) approach: we construct the ``noncommutative Hamiltonian'' corresponding to a simple theta-deformation in order to consider noncommutative corrections to the (formally classical) effective scheme of the flat FLRW LQC (which already carries quantum corrections from LQG).

In view of the above discussion, consider the algebra
\begin{equation}
\{V^{nc},p^{nc}_{\phi}\}=\theta,\ \{\beta^{nc},V^{nc}\}=4\pi G\gamma,\ \{\phi^{nc},p^{nc}_{\phi}\}=1,
\label{ncm-free}
\end{equation}
with the remaining brackets being zero.

The above relations can be implemented by working with the shifted variables
\begin{equation}
V^{nc}=V+a\theta\phi,\ p_{\phi}^{nc}=p_\phi+b\theta\beta,\ \beta^{nc}=\beta,\ \phi^{nc}=\phi,
\label{nc-rel2}
\end{equation}
where $a$ and $b$ satisfy the relation $a-4\pi G\gamma b=1$. Note that $(\beta^{nc},\phi^{nc},V^{nc},p_\phi^{nc})$ denote the non-canonical variables satsifying (\ref{ncm-free}) in the deformed phase space $\tilde\Gamma$. If we were to write the equations of motion employing these non-canonical variables we would have to do so by considering $\tilde{\mathsf{P}}$ in such coordinates. As announced, we prefer to write the equations of motion in standard form by employing appropiate canonical coordinates in $\tilde\Gamma$, which we  denote by $(\beta,\phi,V,p_\phi)$ (the same notation for the canonical coordinates in the undeformed phase space $\Gamma$).

Relations (\ref{nc-rel2}) enables to write the noncommutative Hamiltonian, it takes the form 
\begin{equation}
	\mathcal{H}^{nc}_{eff}=-\frac{3}{8\pi G\gamma^3\lambda^2}\sin^2(\lambda\beta)V^{nc}+\frac{(p_\phi^{nc})^2}{2V^{nc}}+\mathcal{V(\phi)}V^{nc}.\label{ncham}
\end{equation}
The respective noncommutative equations of motion are given by
 \begin{eqnarray}
 \label{feg1} \dot{\beta}=4\pi G\gamma\left[-\frac{3}{4\pi G\gamma^2\lambda^2}\sin^2(\lambda\beta)+2\mathcal{V(\phi)}\right],\\
 \label{feg2} \dot{V}=\frac{3}{\gamma\lambda}V^{nc}\sin(\lambda\beta)\cos(\lambda\beta)-\frac{4\pi G\gamma b\theta p_{\phi}^{nc}}{V^{nc}},\\
 \label{feg3} \dot{\phi}=\frac{p_{\phi}^{nc}}{V^{nc}},\\
 \label{feg4} \dot{p}_{\phi}=\frac{3a\theta}{4\pi G\gamma^{2}\lambda^{2}}\sin^{2}(\lambda\beta)-2a\theta \mathcal{V(\phi)}-V^{nc}\partial_\phi \mathcal{V(\phi)},
 \end{eqnarray}
in the limit $\theta\rightarrow0$ we recover the commutative field equations. From the equation of motion for $\dot\phi$ and $\mathcal{H}_{eff}^{nc}$, we can see that the energy density takes the form $\rho=\dot\phi^2/2+V(\phi)=\rho_c\sin^2(\lambda\beta)$, which is the same functional form (in terms of $\beta$) as in the standard case, and therefore attains the same maximum value $\rho_c$. However, the functional form in terms of $t$ would not be in general the same as the standard case, since the solution of $\beta(t)$ is likely to depend on the noncommutative parameter $\theta$ (nonetheless, given the monotonic behavior of $\beta$, it is expected that the behavior of the energy density function is in the overall the same as in the standard case). It is also noted that several stationary values of the volume function could in principle be attained, which in turn could result in several bouncing scenarios (depending on the evolution of all dynamical variables).

The equation of motion for $V$ indicates that the simultaneous fulfillment of
\begin{equation}
b=0,\ \ V^{nc}(t_c)>0 \label{condb=0}
\end{equation}
(where, as before, $t_c$ is such that $\rho(t_c)=\rho_c$) is a sufficient condition for a minimum volume bounce to be reached when the maximum $\rho_c$ of the energy density $\rho$ is attained (this condition is already present in the free case, but it was not spelled out explicitly in Ref. \cite{hindawi}). That is to say, condition (\ref{condb=0}) implies that the standard LQC Big Bounce is preserved. This will be the only bouncing scenario since $\beta$ is a decreasing function of time (as  (\ref{feg1}) indicates), taking values on $\left(0,\frac{\pi}{\lambda}\right)$, and since $V^{nc}=0$ is not allowed during evolution (in view of the equation of motion for $\phi$ and the energy density being bounded).
	
On the other hand, the simultaneous fulfillment of
\begin{equation}
\dot{\phi}(t_c)=0,\  b\theta\ddot\phi(t_c)<0 \label{condbounce}
\end{equation}
which translates into
\begin{equation}
\rho(\phi(t_c))=\mathcal{V}(\phi(t_c)),\  b\theta\ddot\phi(t_c)<0,
\end{equation}
is also a sufficient condition for the occurrence of a minimum volume bounce; however, it does not precludes the existence of additional bounces (which could lead to additional minima of volume, not necessarily all positive), hence, singularity resolution is not guaranteed to always be achieved in this latter scenario. 
Condition (\ref{condbounce}) would allow to fix the value of $\phi$ at this minimum volume bounce. Furthermore, from the equation of motion for $\phi$ it is observed that $p_\phi(t_c)=-b\theta\beta(t_c)=-\frac{b\theta\pi}{2\lambda}$ must hold. We are facing the following situation: by demanding that a minimum volume bounce takes place when the energy density reaches its maximum value, but not restricting to the particular noncommutativity representation given by $b=0$, initial conditions for all dynamical variables (except for the volume) get fixed at the bounce. This condition therefore leads to a dramatically reduced number of  solutions which feature a bounce somewhat resembling the (single) Big Bounce of standard LQC. This state of affairs is obviously not particularly appealing.

Of course, the conditions (\ref{condb=0}) and (\ref{condbounce}) do not exhaust all the possibilities for the existence of a minimum volume bounce, but they do exhaust all the possibilities for a minimum volume bounce taking place at $\rho=\rho_c$. Additionally, these conditions exclude each other: they cannot both occur for solutions featuring such a bouncing event.

For the scalar field we can construct the modified Klein-Gordon equation using (\ref{feg1})-(\ref{feg4}):
\begin{equation}
\label{mkge}
\ddot\phi+\frac{1}{V^{nc}}\left[\dot V^{nc}+a\theta\phi\right]\dot\phi+\partial_\phi\mathcal{V}(\phi)=0,
\end{equation}
we can see that in the limit $\theta\to0$ we recover the commutative Klein-Gordon equation for the scalar field.  Now we construct the corresponding noncommutative modified Friedmann equation. Since $H=\frac{\dot a}{a}=\frac{\dot V(t)}{3V(t)}$, taking into account the effective Hamiltonian constraint $\mathcal{H}^{nc}_{ eff}\approx0$ and the equation of motion for $V$, we have
\begin{equation}
H^2=\frac{8\pi G}{3}\rho\left[1-\frac{\rho+\rho_\theta}{\rho_c}\right],\label{friedmann-nc}
\end{equation}
where $\rho_\theta$ is the contribution of the noncommutative parameter to the generalization of the modified Friedmann equation. This $\rho_\theta$ can be written as linear order contributions and quadratic contributions in $\theta$, $\rho_\theta=\rho_1(\theta)+\rho_2(\theta^2)$, and these are given as
\begin{eqnarray}
\rho_1(\theta)&=&\frac{2a\theta\phi}{V}(\rho_c-\rho)\nonumber\\
&&+\frac{\sqrt 2b\theta}{\gamma\lambda V}\sqrt{\frac{\rho_{c}}{\rho}\left(1-\frac{\rho}{\rho_c}\right)(\rho-\mathcal{V}(\phi))}\\
\nonumber\rho_2(\theta^2)&=&\frac{a^2\theta^2\phi^2}{V^2}(\rho_c-\rho)\nonumber\\
&&+\frac{\sqrt 2ab\theta^2\phi}{\gamma\lambda V}\sqrt{\frac{\rho_{c}}{\rho}\left(1-\frac{\rho}{\rho_c}\right)(\rho-\mathcal{V}(\phi))}\nonumber\\
&&+\frac{4\pi Gb^2\theta^2}{3V^2}\rho_c\left(1-\frac{\mathcal{V}(\phi)}{\rho}\right).
\end{eqnarray}
From eq. (\ref{friedmann-nc}) it is observed that if condition (\ref{condb=0}) is imposed, the noncommutative contribution is $zero$ at $t=t_c$. On the other hand, for the limited number of bouncing solutions given by imposing condition (\ref{condbounce}), the contribution at $t=t_c$ of terms related to noncommutativity gets also suppressed. 

\section{Noncommutative effective LQC of flat FLRW with scalar field $\phi$ with potential term $\mathcal{V}(\phi)=\frac{1}{2}m^2\phi^2$}
In the following we concentrate on finding particular solutions which fulfill either (\ref{condbounce}) or (\ref{condb=0}) for the case of a quadratic potential term, and on establishing if solutions exist which feature an inflationary scenario with a sufficient number of e-foldings.

When considering the quadratic potential term (\ref{qpotential}),
%
the Hamiltonian (\ref{ncham}), and the equations of motion emanating from it, are given by
\begin{equation}
H^{nc}_{eff}=-\frac{3}{8\pi G\gamma^{2}\lambda^{2}}\sin^{2}(\lambda\beta)V^{nc}+\frac{(p^{nc}_{\phi})^2}{2V^{nc}}+\frac{1}{2}m^2\phi^2V^{nc}\label{ham-const}
\end{equation}
\begin{eqnarray}
\label{betanc}\dot{\beta}&=&-\frac{3}{\gamma\lambda^2}\sin^{2}(\lambda\beta)+4\pi G\gamma m^{2}\phi^{2},\\
\label{volnc}\dot{V}&=&\frac{3}{\gamma\lambda}V^{nc}\sin(\lambda\beta)\cos(\lambda\beta)-4\pi G\gamma b\theta\frac{p_{\phi}^{nc}}{V^{nc}},\\
\label{finc}\dot{\phi}&=&\frac{p^{nc}_{\phi}}{V^{nc}},\\
\label{pfinc}\dot{p}_{\phi}&=&\frac{3a\theta}{8\pi G\gamma^{2}\lambda^{2}}\sin^{2}(\lambda\beta)+a\theta\frac{(p_{\phi}^{nc})^{2}}{2(V^{nc})^{2}}-m^{2}V^{nc}\phi\nonumber\\
&&-\frac{1}{2}a\theta m^{2}\phi^{2},
\end{eqnarray}
Of course, these field equations reduce to the commutative ones when taking $\theta\rightarrow0$, and exhibit the features already noticed in the general case presented in section \ref{section2}. 
\subsection{Particular inflationary solutions related to condition $b=0,\ \ V^{nc}(t_c)>0$}
As we saw earlier, condition (\ref{condb=0}) ensures that all solutions undergo a single minimum volume bounce at $\rho_c$; in other words, a single standard LQC Big Bounce event is always featured for each solution satisfying (\ref{condb=0}). Trajectories with a sufficiently long inflationary phase which pertain to such solutions sector are therefore the best candidates for reproducing an early universe compatible with the (effective) LQC paradigm.

\begin{figure*}[htb]
	\centering
   \subfloat[ ]{\label{rev00}
      \includegraphics[width=.35\textwidth]{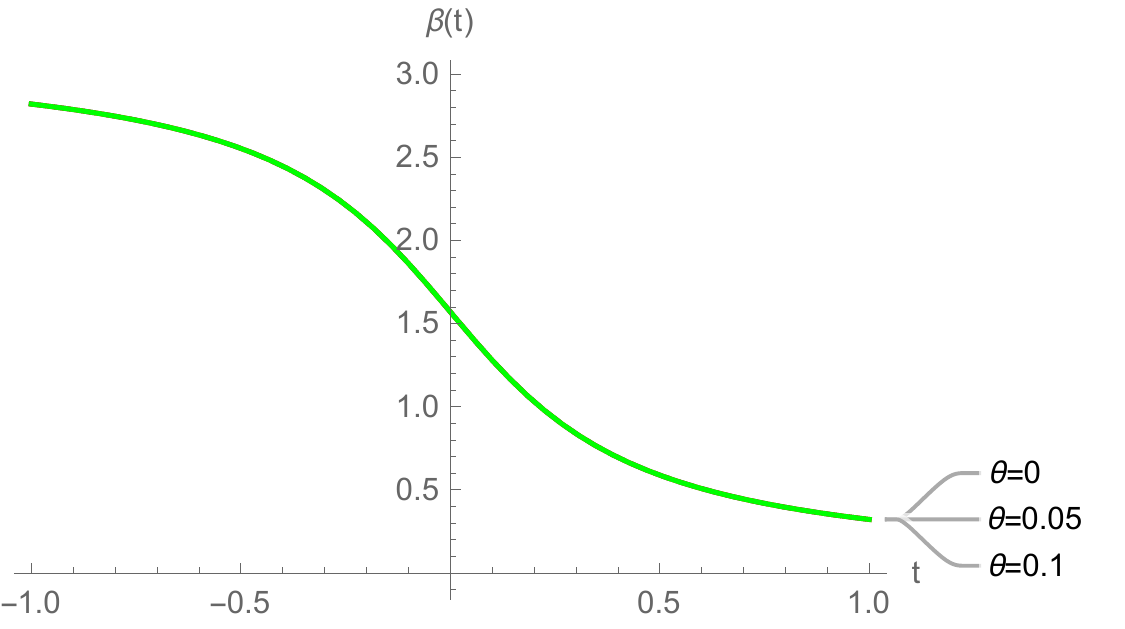}}
~
   \subfloat[ ]{\label{rev_sol01}
      \includegraphics[width=.35\textwidth]{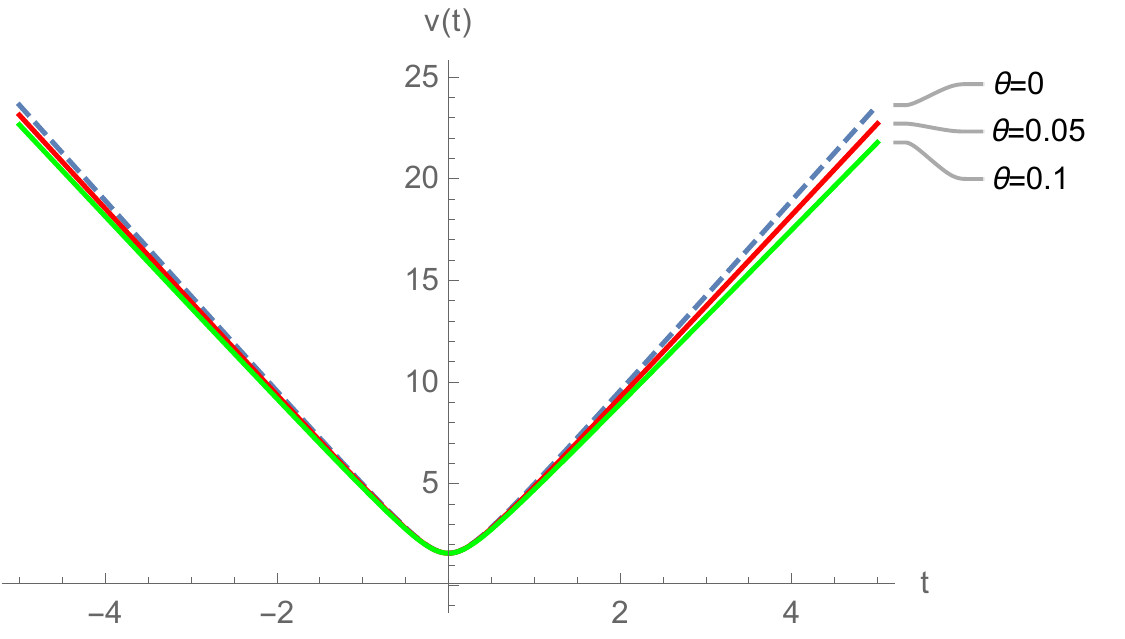}}

   \subfloat[ ]{\label{rev_sol02}
      \includegraphics[width=.35\textwidth]{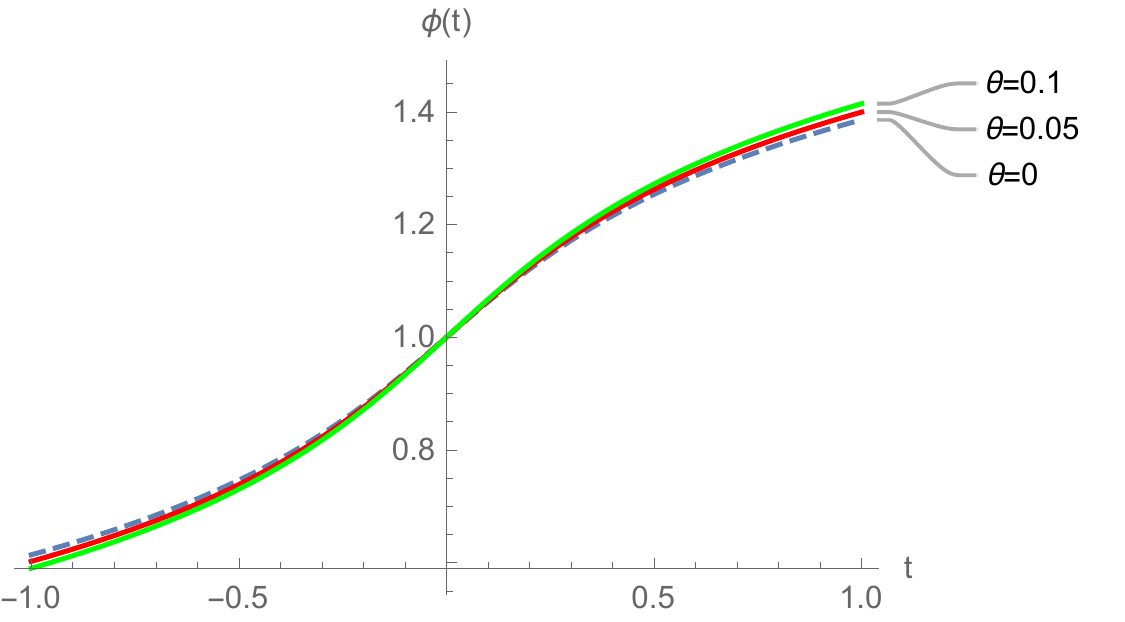}}
~
   \subfloat[ ]{\label{rev_sol03}
      \includegraphics[width=.35\textwidth]{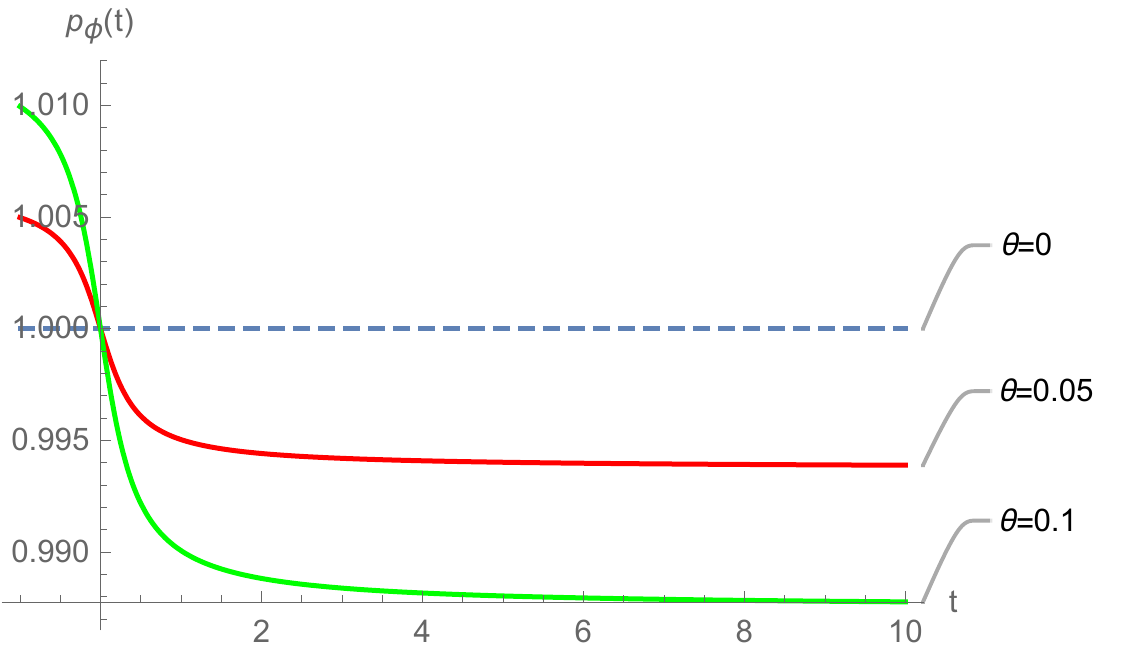}}\\
~      
   \subfloat[ ]{\label{rev_sol04}
      \includegraphics[width=.35\textwidth]{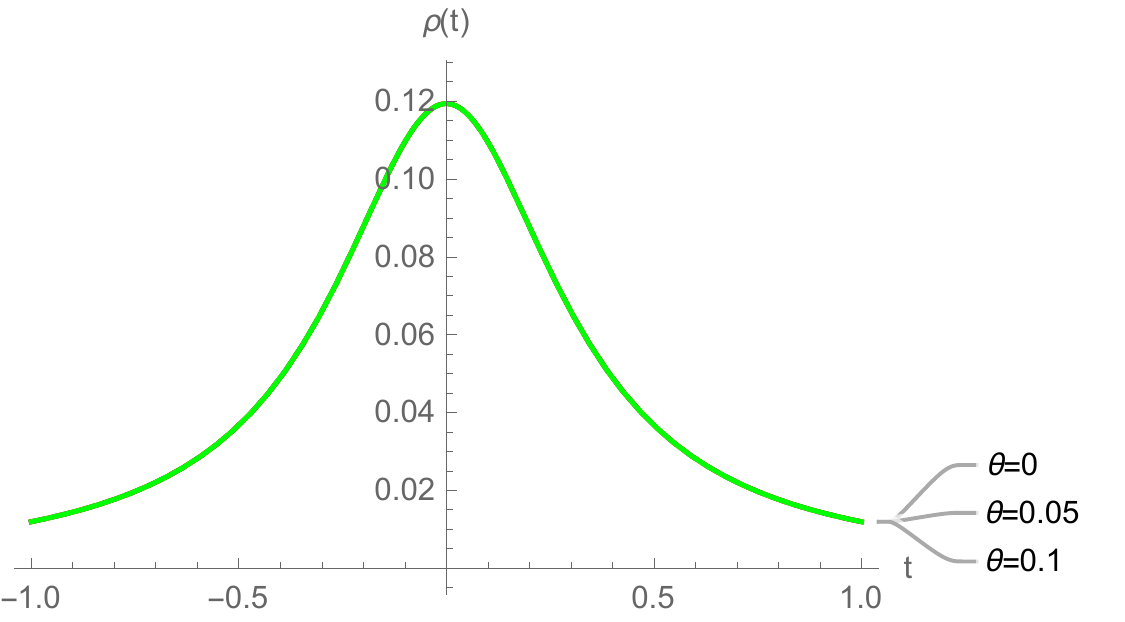}}
\caption{Numerical solutions to (\ref{betanc})-(\ref{pfinc}) taking into account (\ref{condb=0}). The dotted line corresponds to $\theta=0$, green line to $\theta=0.1$ and red line to $\theta=0.05$. (a) $\beta(t)$. (b) volume function, $V(t)$. (c) scalar field, $\phi(t)$. (d) the momentum $p_\phi(t)$ canonically conjugated to the scalar field. (e) energy density, $\rho(t)=\dot\phi^2/2+m^2\phi^2/2$.  The following initial conditions and values were employed: $v(0)=\pi/2\lambda$, $\beta (0)=\pi/2\lambda$, $\phi (0)=1$, $p_\phi(0)=1$, and the constants , $a=-1$, $b=0$, $m=2\times10^{-6}$. In order to better appreciate the differences, we used $\lambda=G=\gamma=1$.}
	\label{sncb0}
\end{figure*}

\begin{figure}[b]
	\begin{center}
		\includegraphics[width=3in]{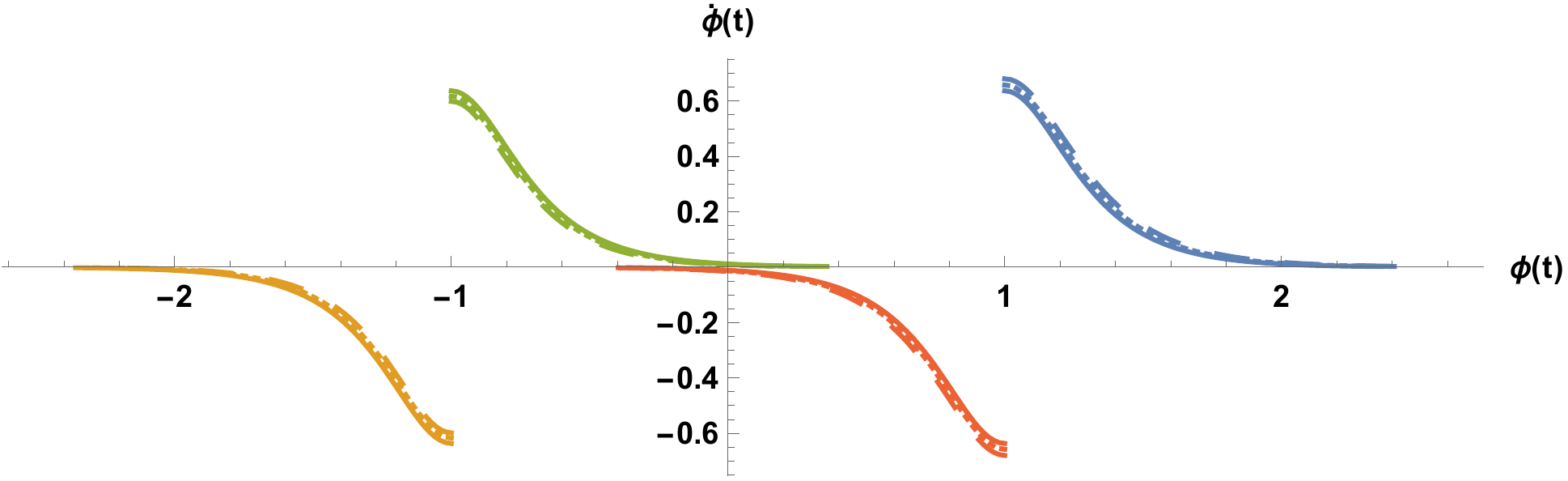}
		\caption{Parametric Plot $(\phi,\dot\phi)$ constructed from numerical solutions to (\ref{betanc})-(\ref{pfinc}) taking into account (\ref{condb=0}). The solid line corresponds to $\theta=0$, the tiny dashed line to $\theta=0.5$ and the large dashed line to $\theta=0.1$. The following initial conditions and values were employed: : $v(0)=\pi/2\lambda$, $\beta (0)=\pi/2\lambda$, $\phi (0)=1$, $p_\phi(0)=1$, and the constants , $a=-1$, $b=0$, $m=2\times10^{-6}$}
		\label{fig:Atb0}
	\end{center}
\end{figure}

Analytical solutions to the equations of motion (\ref{feg1}-\ref{feg4}) could not be found. However, some curves constructed via numerical solutions are shown in Figs. \ref{rev00}-\ref{rev_sol04}. We can explicitly observe that the bounce is preserved, as expected. The energy density, Fig. \ref{rev_sol04}, has the same behavior that the one exhibited in standard LQC. Trajectories in the $\phi-\dot\phi$ plane are presented in Fig. \ref{fig:Atb0}.
\begin{figure}[b]
	\begin{center}
		\includegraphics[width=3.5in]{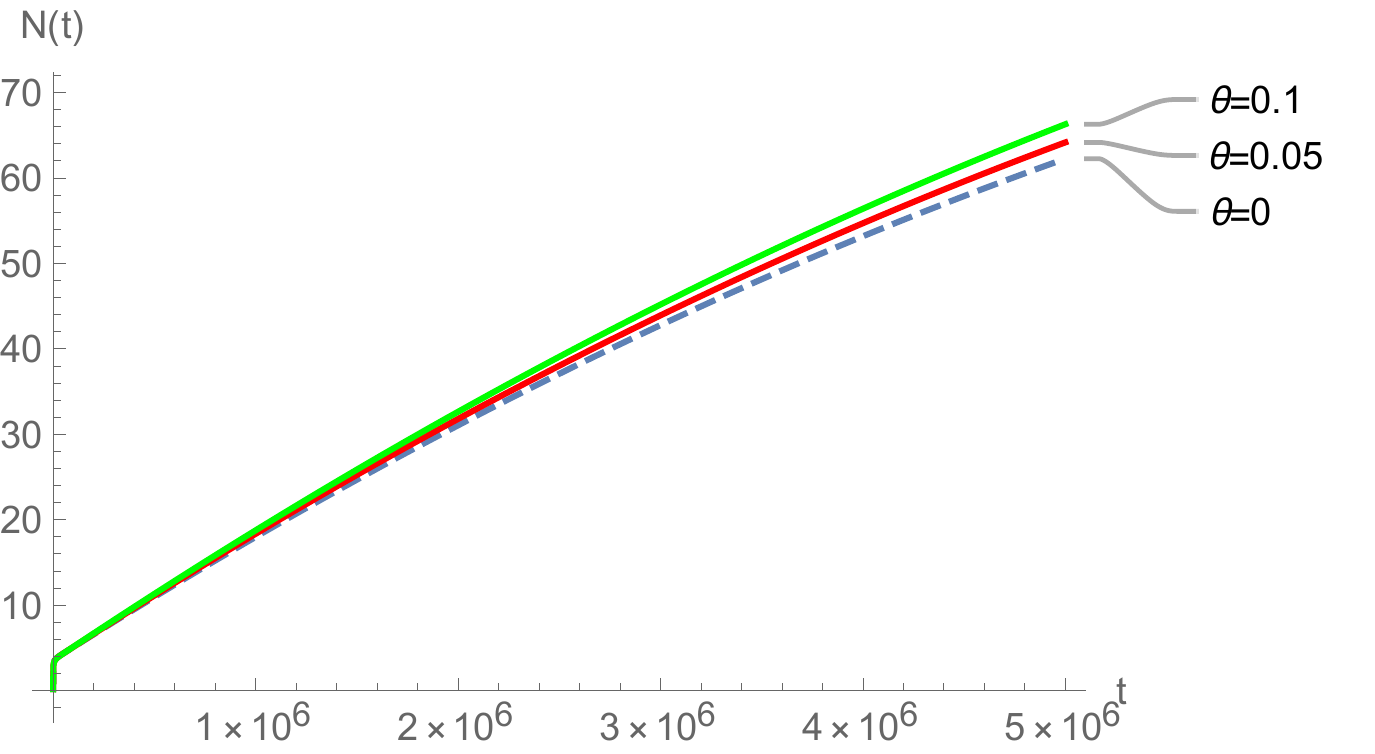}
		\caption{The number $N$ of e-foldings as a function of $t$ constructed from numerical solutions to (\ref{betanc})-(\ref{pfinc}) taking into account (\ref{condb=0}). The dotted line corresponds to $\theta=0$, green line to $\theta=0.1$ and red line to $\theta=0.05$. The following initial conditions and values were employed: $v(0)=\pi/2\lambda$, $\beta (0)=\pi/2\lambda$, $\phi (0)=1$, $p_\phi(0)=1$, and the constants , $a=-1$, $b=0$, $m=2\times10^{-6}$  We observe that a sufficiently large number of e-foldings is achieved.}
		\label{fig:Nb0}
	\end{center}
\end{figure}
%

We can compute the duration of the inflationary period by means of the number $ N $ of e-foldings (this epoch had to last for a long enough period, in order to make the cosmological observations sufficiently flat within the observational limits)
\begin{equation}
N(t)=\int^t_{t_i}H(t)dt
\end{equation}
A period of early inflationary expansion consistent with observations demands $ N \ge 60$ \cite{wmap}. 
Fig. \ref{fig:Nb0} shows the behavior of $N (t)$, where initial conditions associated to (\ref{condb=0}) have been considered. We observe that the model presents an early inflationary period with $N\ge60$. 

\subsection{Particular inflationary solutions related to condition $\dot{\phi}(t_c)=0,\ b\theta\ddot\phi(t_c)<0$}
In this case, condition (\ref{condbounce}) takes the following form
\begin{equation}
\label{condbouncephi2}
p_{\phi}(0)=-\frac{b\theta\pi}{2\lambda},\ b\theta\ddot\phi(t_c)<0,\ \phi(0)=\left\{\begin{matrix}\sqrt{\frac{2\rho_c}{m^2}},&b\theta>0\\-\sqrt{\frac{2\rho_c}{m^2}},&b\theta<0\end{matrix}\right..
\end{equation}

In other words, under this assumption, the standard bounce takes place only for the solutions related to the above initial conditions which correspond to only \textit{one} trajectory on the $\phi-\dot{\phi}$ plane for each case $b\theta>0$, $b\theta<0$.

\begin{figure}[h]
	\begin{center}
		\includegraphics[width=3in]{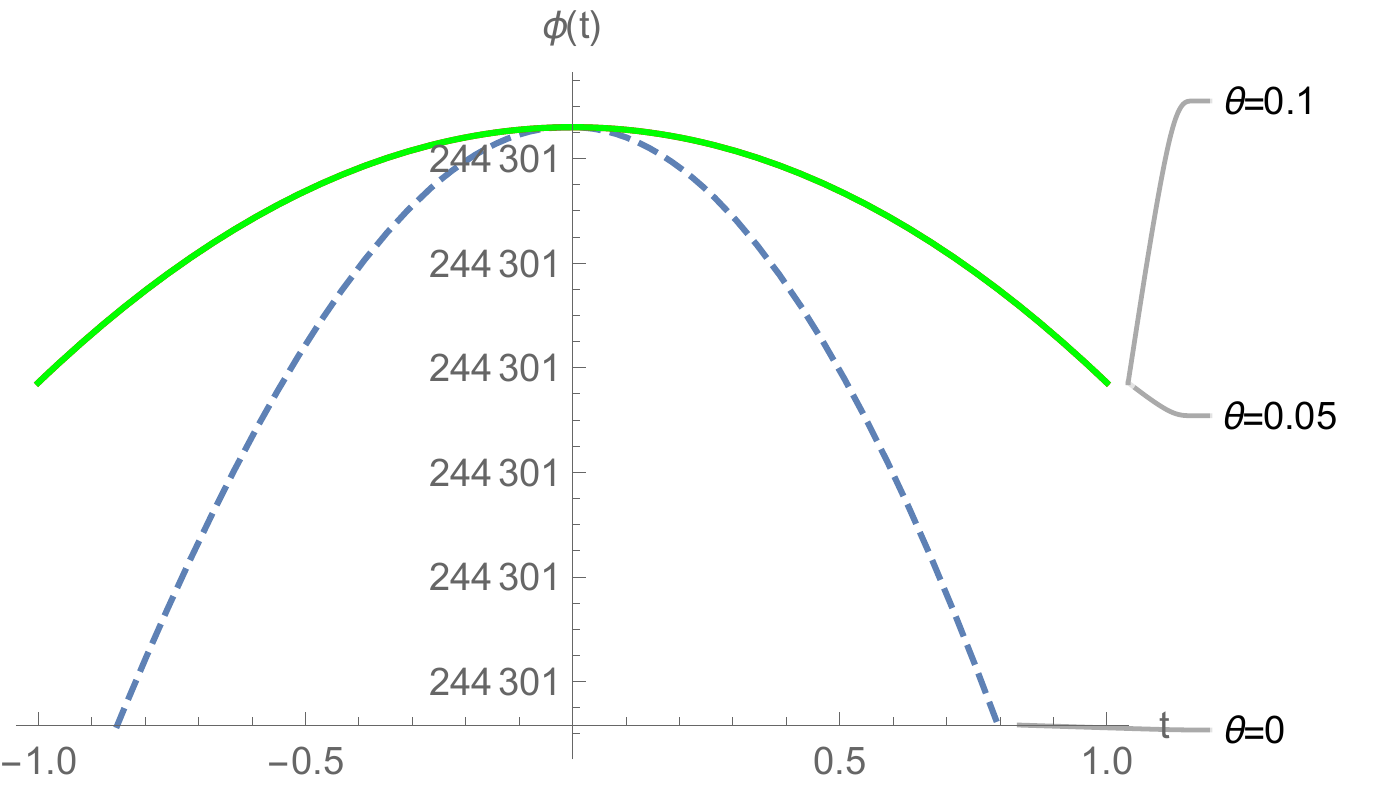}
		\caption{Numerical solution for (\ref{finc}) taking into account (\ref{condbouncephi2}). The dotted line corresponds to $\theta=0$, green line to $\theta=0.1$ and red line to $\theta=0.05$. The following initial conditions and values were employed: $\beta(0)=\pi/2\lambda$, $V(0)=\pi/2\lambda$, $\phi(0)=\sqrt{2\rho_c/m^2}$, $p_\phi(0)=-(b\theta\pi)/2\lambda$; $\lambda=G=\gamma=1$, $b=1$, $m=2\times10^{-6}$.}
		\label{fig:phigen}
	\end{center}
\end{figure}

Condition (\ref{condbouncephi2}) implies that, in the case $b\theta>0$, a local maximum of the scalar field $\phi (t)$ is reached at $ t_c $, for any value of the $\theta$ parameter. This can be seen represented in the numerical solution shown in Fig. \ref{fig:phigen}.
%

\begin{figure}[b]
	\centering
   \subfloat[ ]{\label{fig:Vgen0}
      \includegraphics[width=.5\textwidth]{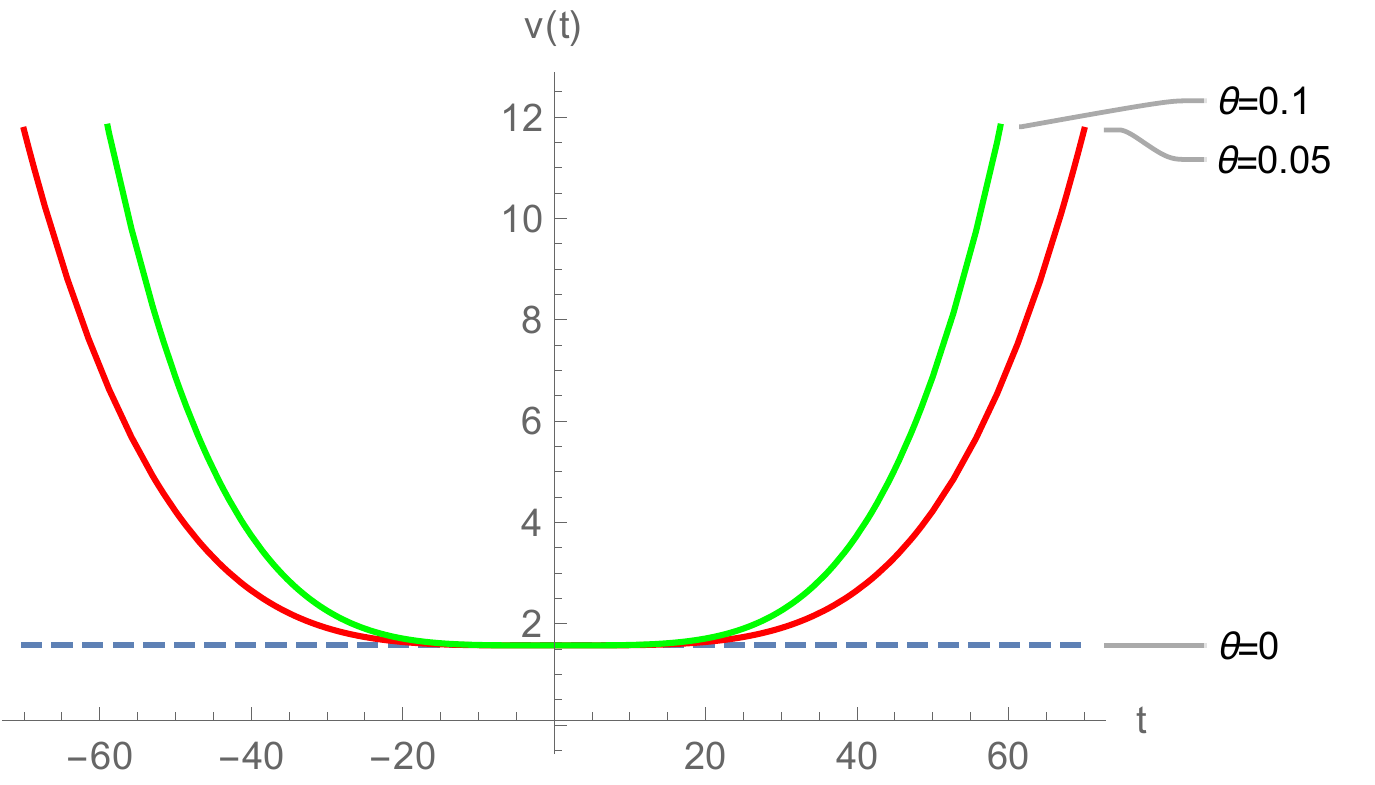}}

      %
	\caption{Numerical solution to (\ref{volnc}) taking into account \ref{condbouncephi2}. The dotted line corresponds to $\theta=0$, green line to $\theta=0.1$ and red line to $\theta=0.05$. The following initial conditions and values were employed: $\beta(0)=\pi/2\lambda$, $V(0)=\pi/2\lambda$, $\phi(0)=\sqrt{2\rho_c/m^2}$, $p_\phi(0)=-(b\theta\pi)/2\lambda$; $\lambda=G=\gamma=1$, $b=1$, $m=2\times10^{-6}$.}
	\label{fig:Vgen}
\end{figure}

\begin{figure}[t]
	\centering
   \subfloat[ ]{\label{fig:betagen}
      \includegraphics[width=.4\textwidth]{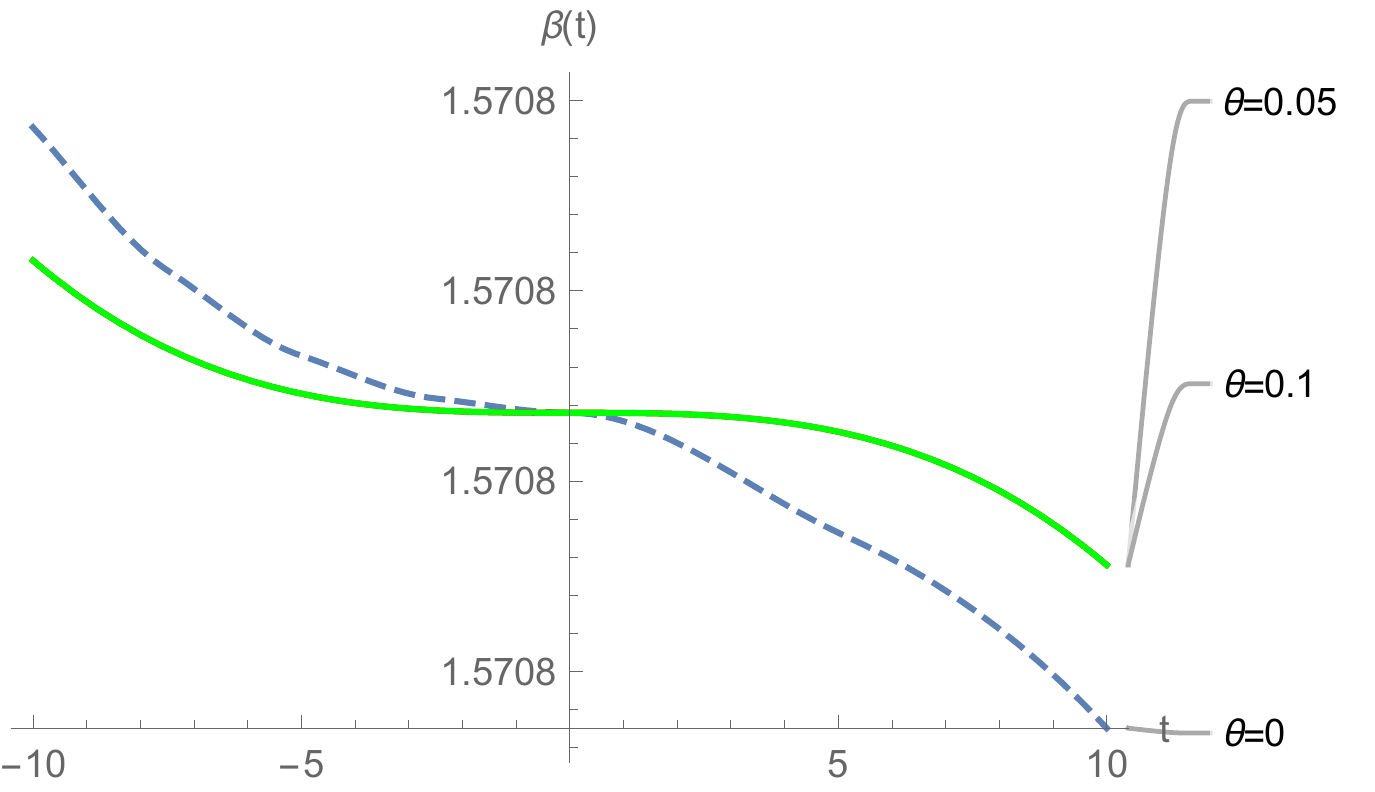}}
\\
~
   \subfloat[ ]{\label{fig:dengen0}
      \includegraphics[width=.4\textwidth]{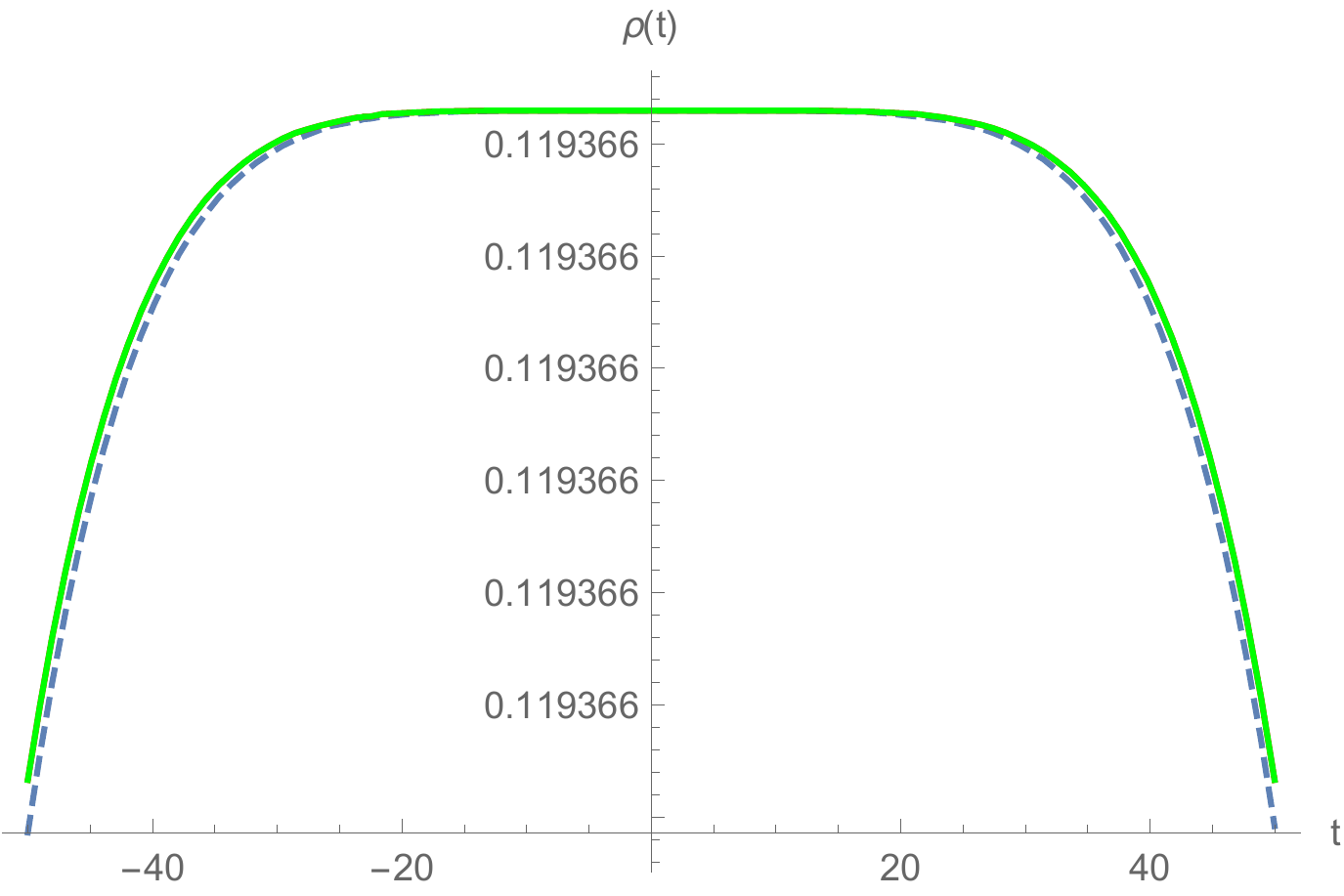}}
      
	\caption{Numerical solutions to (\ref{betanc})-(\ref{pfinc}) taking into account \ref{condbouncephi2}. The dotted line corresponds to $\theta=0$, green line to $\theta=0.1$ and red line to $\theta=0.05$. (a): $\beta(t)$; (b): energy density function (\ref{densitync}). The following initial conditions and values were employed: $\beta(0)=\pi/2\lambda$, $V(0)=\pi/2\lambda$, $\phi(0)=\sqrt{2\rho_c/m^2}$, $p_\phi(0)=-(b\theta\pi)/2\lambda$; $\lambda=G=\gamma=1$, $b=1$, $m=2\times10^{-6}$.}
	\label{fig:dengen}
\end{figure}

\begin{figure}[t]
	\centering
   \subfloat[ ]{\label{Nnc_sol01}
      \includegraphics[width=.5\textwidth]{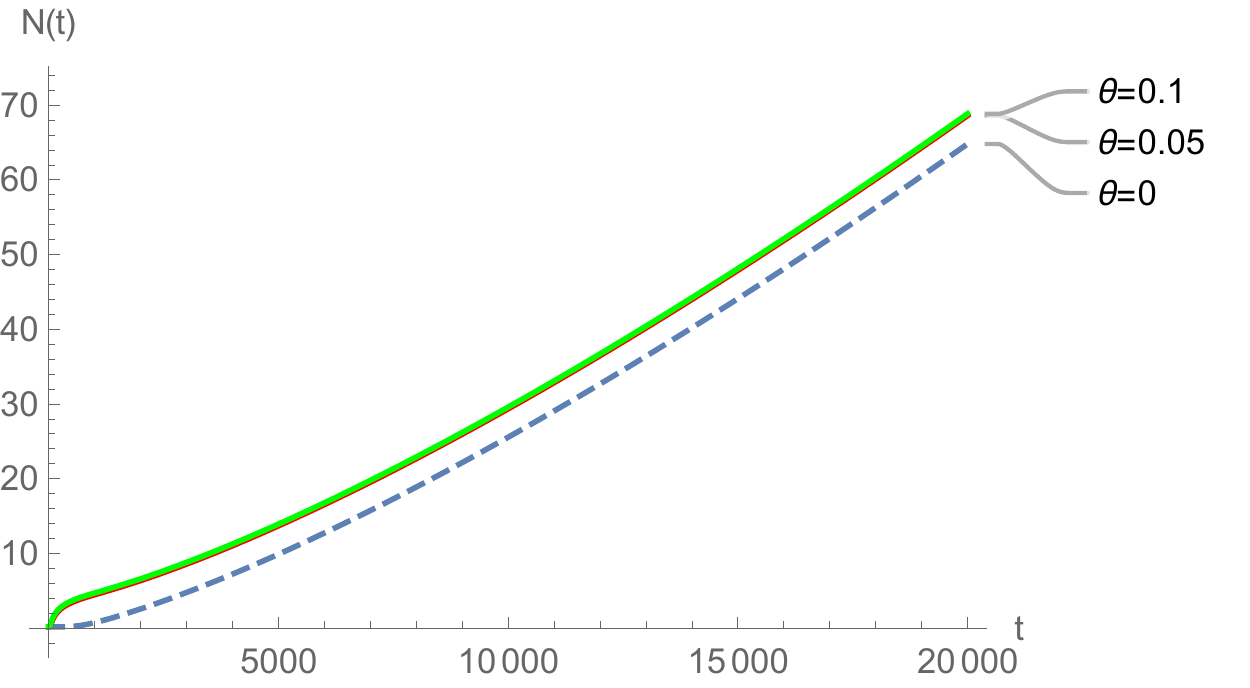}}

   \subfloat[ ]{\label{Nnc_sol02}
      \includegraphics[width=.4\textwidth]{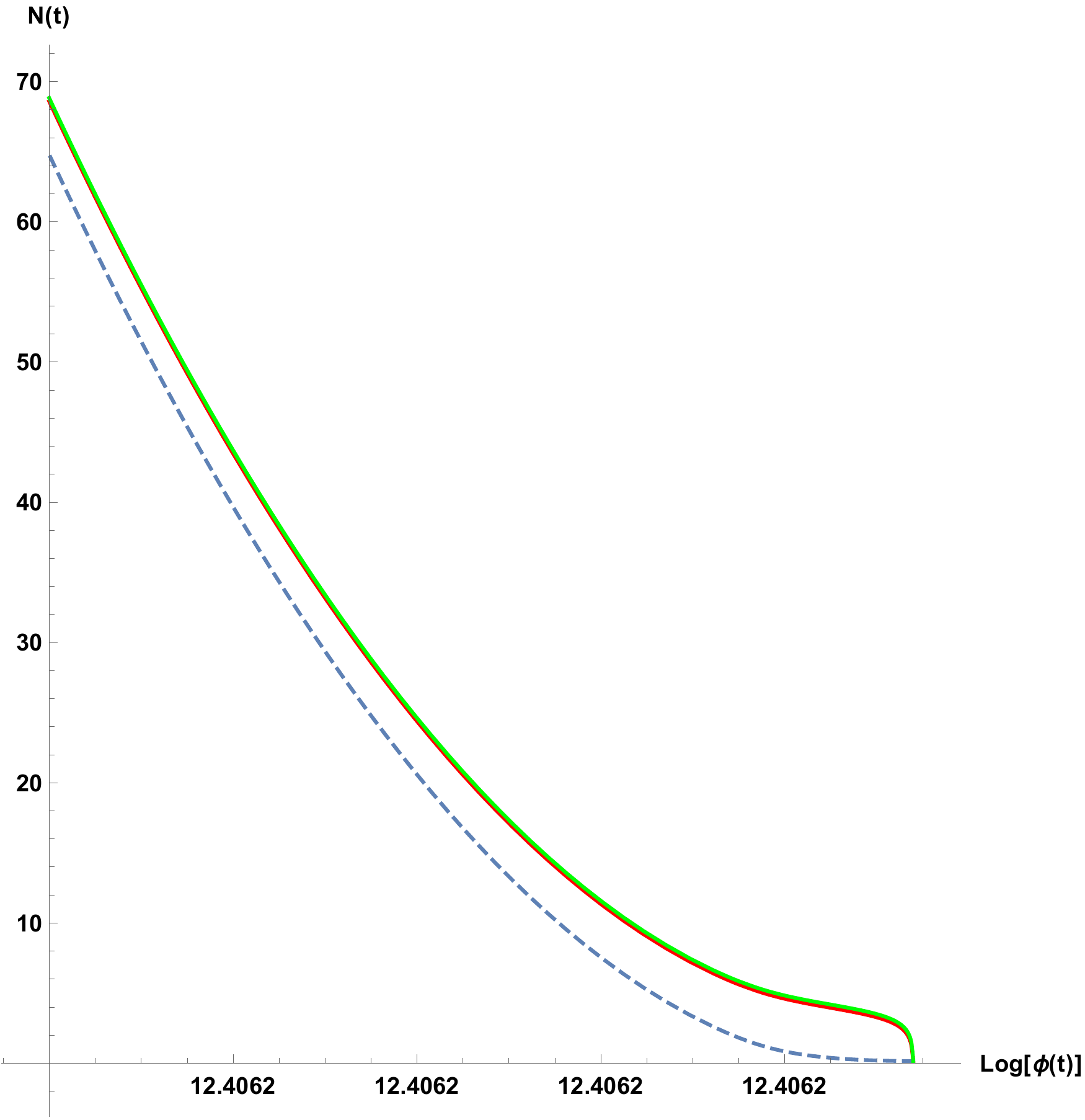}}

	\caption{The number $N$ of e-foldings as a function of $t$ constructed from numerical solutions to (\ref{betanc})-(\ref{pfinc}) taking into account (\ref{condbouncephi2}). The dotted line corresponds to $\theta=0$, green line to $\theta=0.1$ and red line to $\theta=0.05$. The following initial conditions and values were employed: $\beta(0)=\pi/2\lambda$, $V(0)=\pi/2\lambda$, $\phi(0)=\sqrt{2\rho_c/m^2}$, $p_\phi(0)=-(b\theta\pi)/2\lambda$; $\lambda=G=\gamma=1$, $b=1$, $m=2\times10^{-6}$.}
	\label{fig:Ngen}
\end{figure}

From Fig. \ref{fig:Vgen} we observe that the inclusion of noncommutativity results in a greater volume expansion rate.
The energy density 
$\rho=\frac{\dot{\phi}^{2}}{2}+\frac{1}{2}m^2\phi^{2}$ is given by
\begin{equation}
\label{densitync}
\rho=\frac{1}{2}\left(\frac{p_{\phi}^{nc}}{V^{nc}}\right)^{2}+\frac{1}{2}m^{2}\phi^{2}=\frac{3}{8\pi G\gamma^{2}\lambda^{2}}\sin^{2}(\lambda\beta),
\end{equation}
In Fig. \ref{fig:dengen0} the energy density is shown, it is observed that it has a dependence on the noncommutative parameter, in contrast to the free case \cite{ijmpd}. As already pointed out, this is because the energy density (\ref{densitync}) depends explicitly on $\beta(t)$, but $\beta$ is $\theta$-dependent, as shown in Figure (\ref{fig:betagen}). It is important to emphasize that although the energy density depends on the noncommutative parameter, the maximum $\rho_{c}$ is conserved for any value of $\theta$ (this was also already pointed out when considering the general case in the preceding section). 
As in the previous subsection, we can establish if an inflationary period with enough e-foldings is achieved. \ref{Nnc_sol01} shows $N$ as a function of $t$, taking initial conditions consistent with (\ref{condbouncephi2}). The model presents a sufficiently long period of inflation. On the other hand Fig.  \ref{Nnc_sol02} returns the number $N$ of e-foldings in terms of the scalar field $\phi$. The value employed for the field is $\phi\approx10^6$, which is determined from $\mathcal V(\phi(t_c))=\rho_c$.
%
Finally, we spell out the noncommutative modified Friedmann equation (\ref{friedmann-nc}) specialized to the quadratic potential term given in (\ref{qpotential}): 
\begin{equation}
\label{mfep2}
H^2=\frac{8\pi G}{3}\rho\left[1-\frac{\rho+\rho_\theta}{\rho_c}\right],
\end{equation}
where $\rho^\theta=\rho_1(\theta)+\rho_2(\theta^2)$ and
\begin{eqnarray}
\rho_1(\theta)&=&\frac{2a\theta\phi}{V}(\rho_c-\rho)\nonumber\\
&&+\frac{\sqrt 2b\theta}{\gamma\lambda V}\sqrt{\frac{\rho_{c}}{\rho}\left(1-\frac{\rho}{\rho_c}\right)(\rho-\frac{1}{2} m^2\phi^2)}\\
\nonumber\rho_2(\theta^2)&=&\frac{a^2\theta^2\phi^2}{V^2}(\rho_c-\rho)\\
&&+\frac{\sqrt 2ab\theta^2\phi}{\gamma\lambda V}\sqrt{\frac{\rho_{c}}{\rho}\left(1-\frac{\rho}{\rho_c}\right)(\rho-\frac{1}{2} m^2\phi^2)}\nonumber\\
&&+\frac{4\pi Gb^2\theta^2}{3V^2}\rho_c\left(1-\frac{ m^2\phi^2}{2\rho}\right).
\end{eqnarray}
This equation gives us the dynamical evolution of the model. If we take the limit $\theta\to 0$, we recover the usual modified Friedmann equation of standard LQC.
 
\section{Discussion}
The present investigation shows that the inclusion of a generic potential term in the momentum sector noncommutativity encoded in the algebra (\ref{ncm-free}), within the LQC effective scheme of the flat FLRW, maintains the occurrence of the single Big Bounce scenario characteristic of standard LQC (considering the class of noncommutative models associated to condition (\ref{condb=0})). The alternative relations (\ref{condbounce}) were established as also a sufficient condition for retaining a Big Bounce, however, such condition does not precludes additional stationary points in the volume function (in contrast to condition (\ref{condb=0})) which could lead to several bouncing scenarios. A solution featuring a minimum volume bounce at $\rho=\rho_c$ cannot fulfill both (\ref{condb=0}) and (\ref{condbounce}). As a first approach to studying the inflationary scenario, we specialized to a quadratic potential term (\ref{qpotential}). In this particular case, condition (\ref{condbounce}) reduces to (\ref{condbouncephi2}) which, for $b\theta>0$ ($b\theta<0$), is equivalent to demanding a maximum (minimum) of the scalar field at the bounce; this in turn fixes initial conditions at the bounce for the scalar field and its conjugate momentum: fine tuning reduces to volume-tuning (since $\beta$ is already fixed at the bounce). Figures \ref{fig:dengen}, \ref{fig:Ngen} show the behavior of a particular set of solutions satisfying initial conditions consistent with (\ref{condbouncephi2}), it is observed that a sufficiently large period of inflation is achieved. On the other hand, a satisfactory inflationary period is also achieved for particular sets of solutions consistent with condition (\ref{condb=0}). One of such sets is considered in Figs. \ref{rev00}-\ref{rev_sol04}; \ref{fig:Nb0}. We therefore conclude that solutions to the noncommutative paradigm encoded in (\ref{ncm-free}) exist, which depict an early universe whose behavior is essentially the one emanating from the standard LQC paradigm. There remains to be addressed the question of wether the sufficiently long inflationary period exhibited in the specific solutions found is a rather generic feature (in the sense of  Ref. \cite{sloan}). 
Also, it is important to note that since conditions (\ref{condb=0}) and (\ref{condbounce}) were established by considering a generic potential term $\mathcal V$, the study of the occurrence of a more sensible inflationary scenario by considering other meaningful potential terms is therefore a natural pursuit. These and other related matters are currently being tackled by the authors and will be reported elsewhere.


\section*{Acknowledgments}
J.S. was partially funded by PRODEP grant UGTO-CA-3. A.E.G. was partially supported by a CONACyT PhD fellowship.


\end{document}